\documentclass[10pt]{article}
\usepackage[square,authoryear]{natbib}
\usepackage{graphicx}
\usepackage{marsden_article}
\usepackage{framed}
\usepackage{epstopdf,framed}
\usepackage{pdfsync}
\usepackage{latexsym,amsmath,amscd,amssymb,framed,graphics,color,
mathrsfs,stmaryrd}
\usepackage{eucal}
\begin{document}
\newtheorem{remark}[theorem]{Remark}

\title{From Lagrangian mechanics to nonequilibrium thermodynamics: a variational perspective}
\vspace{-0.2in}

\newcommand{\todoFGB}[1]{\vspace{5 mm}\par \noindent
\framebox{\begin{minipage}[c]{0.95 \textwidth} \color{red}FGB: \tt #1
\end{minipage}}\vspace{5 mm}\par}


\author{\hspace{-1cm}
\begin{tabular}{cc}
Fran\c{c}ois Gay-Balmaz &
Hiroaki Yoshimura
\\ CNRS - Ecole Normale Sup\'erieure & School of Science and Engineering
\\ Laboratoire de m\'et\'eorologie dynamique  & Waseda University
\\  24 Rue Lhomond 75005 Paris, France & Okubo, Shinjuku, Tokyo 169-8555, Japan
\\francois.gay-balmaz@lmd.ens.fr & yoshimura@waseda.jp\\
\end{tabular}\\\\
}
\date{}
\maketitle
\vspace{-0.3in}

\begin{center}
\abstract{In this paper, we survey our recent results on the variational formulation of nonequilibrium thermodynamics for the finite dimensional case of discrete systems as well as for the infinite dimensional case of continuum systems. Starting with the fundamental variational principle of classical mechanics, namely, Hamilton's principle, we show, with the help of thermodynamic systems with gradually increasing level complexity, how to systematically extend it to include irreversible processes. In the finite dimensional cases, we treat systems experiencing the irreversible processes of mechanical friction, heat and mass transfer, both in the adiabatically closed and in the open cases. On the continuum side, we illustrate our theory with the example of  multicomponent Navier-Stokes-Fourier systems.}
\vspace{2mm}

\end{center}
\tableofcontents

\section{Introduction}

This paper makes a review of our recent results on the development of a variational formulation of nonequilibrium thermodynamics, as established in \cite{GBYo2017a, GBYo2017b,GBYo2018a, GBYo2018b}. This formulation is an extension to nonequilibrium thermodynamics of the Lagrangian formulation of classical and continuum mechanics which includes irreversible processes such as friction, heat and mass transfer, chemical reactions, and viscosity.

\paragraph{Some history of the variational approaches to thermodynamics.} 
Thermodynamics was first developed to treat exclusively equilibrium states and the transition from one equilibrium state to another in which change in temperature plays an important role. In this context, thermodynamics appeared mainly as a theory of heat and is viewed today as a branch of \textit{equilibrium thermodynamics}. Such a classical theory, \textit{which does not aim to describe the time evolution} of the system, can be developed in a well-established setting (\cite{Gibbs1902}), governed by the well-known first and second laws, e.g., \cite{Ca1985}, \cite{LaLi1969_5}. It is worth noting that classical mechanics, fluid dynamics and electromagnetism, being essentially dynamical theories, \textit{cannot} be treated in the context of equilibrium thermodynamics. Although much effort has been done in the theoretical investigation of nonequilibrium thermodynamics in relation with physics, chemistry, biology and engineering, the theory of nonequilibrium thermodynamics has not reached a level of completeness. This is in part due to the lack of a general variational formulation for nonequilibrium thermodynamics that would reduce to the classical Lagrangian variational formulation of mechanics in absence of irreversible processes. So far, various variational approaches have been proposed in relation with nonequilibrium thermodynamics such as the \textit{principle of least dissipation of energy} introduced in \cite{Onsager1931}  later extended in \cite{OnMa1953} and \cite{MaOn1953} that underlies the reciprocal relations in linear phenomenological laws, and the \textit{principle of minimum entropy production} by \cite{Prigogine1947} and \cite{GlPr1971} that gives conditions on steady state processes. Onsager's approach was generalized in \cite{Ziegler1968} to the case of systems with nonlinear phenomenological laws. We refer to \cite{Gyarmati1970} for reviews and developments of Onsager's variational principles, and for a study of the relation between Onsager's and Prigogine's principles. We also refer to \cite[\S6]{Lavenda1978} and \cite{Ichiyanagi1994} for overviews on variational approaches to irreversible processes.  Note that, however, the variational principles developed in these past works are not natural extensions of Hamilton's principle of classical mechanics, because they do not recover Hamilton's principle for the case in which irreversible processes are not included. Another important work  was done by \cite{Bi1975, Bi1984} in conjunction with thermoelasticity, viscoelasticity and heat transfer, where a \textit{principle of virtual dissipation} in a generalized form of d'Alembert principle was used with various applications to nonlinear irreversible thermodynamics. In particular, \cite{Bi1975} mentioned that the relations between the rate of entropy production and state variables may be given as \textit{nonholonomic constraints}. Nevertheless, this variational approach was restricted to \textit{weakly irreversible systems} or \textit{thermodynamically holonomic and quasi-holonomic}. More recently, it was noteworthy that  \cite{FuFu2012} showed a variational formulation for viscoelastic fluids, in which the internal conversion of mechanical power into heat power due to frictional forces was written as a nonholonomic constraint. However, it should be noted that the approaches mentioned above do not present a  systematic and general variational formulation of nonequilibrium thermodynamics and hence are restricted to some class of thermodynamic systems.

The geometry of \textit{equilibrium} thermodynamics has been mainly studied via contact geometry, following the initial works of \cite{Gibbs1873a,Gibbs1873b} and \cite{Ca1909}, by \cite{He1973} and further developments by \cite{Mr1978,Mr1980,MrNuScSa1991}. In this geometric setting, thermodynamic properties are encoded by Legendre submanifolds of the thermodynamic phase space. A step towards a geometric formulation of irreversible processes was made in \cite{EbMaVa2007} by lifting port Hamiltonian systems to the thermodynamic phase space. The underlying geometric structure in this construction is again a contact form. A description of irreversible processes using modifications of Poisson brackets has been initiated in \cite{Gr1984,Ka1984,Mo1984}. This has been further developed for instance in \cite{EdBe1991a,EdBe1991b} and \cite{Mo1986,GrOt1997,OtGr1997}. A systematic construction of such brackets from the variational formulation given in the present paper was presented in \cite{ElGB2018} for the thermodynamics of multicomponent fluids.

\paragraph{Main features of our variational formulation.} The variational formulation for nonequilibrium thermodynamics developed in \cite{GBYo2017a, GBYo2017b,GBYo2018a, GBYo2018b} is distinct from the earlier variational approaches mentioned above, both in its physical meaning and in its mathematical structure, as well as in its goal. Roughly speaking, while most of the earlier variational approaches mainly underlie the equation for the rate of entropy production, in order to justify the expression of the phenomenological laws governing the irreversible processes involved, our variational approach aims to underlie the {\it complete set of time evolution equations of the system\/}, in such a way that it extends the classical Lagrangian formulation in mechanics to nonequilibrium thermodynamic systems including irreversible processes.

This is accomplished by constructing a generalization of the {\it Lagrange-d'Alembert principle\/} of nonholonomic mechanics, where the entropy production of the system, written as the sum of the contribution of each of the irreversible processes, is incorporated into a {\it nonlinear nonholonomic constraint\/}. As a consequence, all the {\it phenomenological laws\/} are encoded in the nonlinear  nonholonomic constraints, to which we naturally associate a {\it variational constraint\/} on the allowed variations of the action functional. A natural definition of the variational constraint in terms of the phenomenological constraint is possible thanks to the introduction of the concept of \textit{thermodynamic displacement}, which generalizes the concept of thermal displacement given by \cite{GrNa1991} to all the irreversible processes. 

More concretely, if the system involves internal irreversible processes, denoted $\alpha$, and irreversible process at the ports, denoted $\beta$, with thermodynamic fluxes $J_\alpha, J_\beta$ and thermodynamic affinities $X^\alpha, X^\beta$ together with a thermodynamic affinity $X^\beta_{\rm ext}$ associated with the exterior, then the thermodynamic displacements $ \Lambda^\alpha, \Lambda^\beta$ are such that $\dot\Lambda  ^\alpha=X^ \alpha$ and $\dot\Lambda  ^\beta=X^ \beta$. This allows us to formulate the variational constraint associated to the phenomenological constraint in a systematic way, namely, by replacing all the velocities by their corresponding virtual displacement and by removing the external thermodynamic affinity $X^\beta_{\rm ext}$ at the exterior of the system as follows:
\[
J_ \alpha \dot \Lambda  ^\alpha +J_ \beta \big(\dot \Lambda  ^\beta - X^\beta_{\rm ext}\big) \;\;\leadsto \;\; J_ \alpha  \delta \Lambda  ^\alpha + J_ \beta  \delta \Lambda  ^ \beta.
\]
Our variational formulation has thus a clear and systematic structure that appears to be common for the macroscopic description of the nonequilibrium thermodynamics of physical systems.  It can be applied to the finite dimensional case of discrete systems such as classical mechanics, electric circuits, chemical reactions, and mass transfer. Further, our variational approach can be naturally extended to the infinite dimensional case of continuum systems; for instance, it can be applied to some nontrivial example such as the {\it multicomponent Navier-Stokes-Fourier equations}. Again, it is emphasized that our variational formulation consistently recovers Hamilton's principle in classical mechanics when irreversible processes are not taken into account.

\paragraph{Organization of the paper.} In \S\ref{sect_VFLM}, we start with a very elementary review of Hamilton's variational principle in classical mechanics and its extension to the case of mechanical systems with external forces. We also make a brief review of the variational formulation of mechanical systems with linear nonholonomic constraints by using the Lagrange-d'Alembert principle. Furthermore, we review the extension of Hamilton's principle to continuum systems and illustrate it with the example of compressible fluids in the Lagrangian description. The variational principle in the Eulerian description is then deduced in the context of {\it symmetry reduction}.  In \S\ref{sec_finite}, we recall the two laws of thermodynamics as formulated by \cite{StSc1974}, and we present the variational formulation of nonequilibrium thermodynamics for the finite dimensional case of discrete systems. We first consider adiabatically closed simple systems and illustrate the variational formulation with the case of a movable piston containing an ideal gas and with a system consisting of a chemical species experiencing diffusion between several compartments. We then consider adiabatically closed non-simple systems such as the adiabatic piston with two cylinders and a system with a chemical species experiencing both diffusion and heat conduction between two compartments. Further we consider the variational formulation for open systems and illustrate it with the example of a piston device with ports and heat sources. In \S\ref{sec_continuum}, we extend the variational formulation of nonequilibrium thermodynamics to the infinite dimensional case of continuum systems and consider a multicomponent compressible fluid subject to the irreversible processes due to viscosity, heat conduction, and diffusion. The variational formulation is first given in the Lagrangian description, from which the variational formulation in the Eulerian description is deduced. This is illustrated with the multicomponent Navier-Stokes-Fourier equations. In \S\ref{conclusions}, we make some concluding remarks and mention further developments based on the variational formulation of nonequilibrium thermodynamics such as variational discretizations, Dirac structures in thermodynamics, reduction by symmetries, and thermodynamically consistent modeling.

\section{Variational principles in Lagrangian mechanics}\label{sect_VFLM}

\subsection{Classical mechanics}\label{sec_LM}

One of the most fundamental statement in classical mechanics is the principle of critical action or Hamilton's principle, according to which the motion of a mechanical system between two given positions is given by a curve that makes the integral of the Lagrangian of the system to be critical (see, for instance, \cite{LaLi1969_1}).

Let us consider a mechanical system with configuration manifold $Q$. For instance, for a system of $N$ particles moving in the Euclidean $3$-space, the configuration manifold is $Q=\mathbb{R}^{3N}$, whereas for a rigid body moving freely in space, $Q=\mathbb{R}^3\times SO(3)$, the product of the Euclidean $3$-space and the rotation group.
Let us denote by $(q^1,...,q^n)$ the local coordinates of the manifold $Q$, also known as generalized coordinates of the mechanical system. Let $L$ be a given Lagrangian of the system, which usually depends only on the position $q$ and velocity $v$ of the system and is hence defined on the tangent bundle\footnote{The \textit{tangent bundle} of a manifold $Q$ is the manifold $TQ$ given by the collection of all tangent vectors in $Q$. As a set it is given by the disjoint union of the \textit{tangent spaces} of $Q$, that is, $TQ=\sqcup_{q\in Q}T_qQ$, where $T_qQ$ is the tangent space to $Q$ at $q$. The elements in $T_qQ$ are denoted $(q,v)$.}, or {\it velocity phase space}, $TQ$ of the manifold $Q$. The Lagrangian $L$ is usually given by the kinetic minus the potential energy of the system as $L(q, v)= K(q, v)-U(q)$.

Hamilton's principle is written as follows. Suppose that the system occupies  the positions $q_1$ and $q_2$ at the time $t_1$ and $t_2$, then the motion $q(t)$ of the mechanical system between these two positions is a solution of the critical point condition
\begin{equation}\label{HP_epsilon}
\left.\frac{d}{d\epsilon}\right|_{\epsilon =0}  \int_{t_1}^{t_2} L\big(q(t,\epsilon), \dot{q}(t, \epsilon)\big){\rm d}t=0,
\end{equation}
where $q(t, \epsilon)$, $t \in [t_{1},t_{2}]$, $\epsilon\in [-a,a]$, is an arbitrary variation of the curve $q(t)$ with fixed endpoints, i.e., $q(t, \epsilon)|_{\epsilon=0}=q(t)$ and $q(t_1, \epsilon)=q(t_1)$, $q(t_2, \epsilon)=q(t_2)$, for all $\epsilon$. The infinitesimal variation associated to a given variation $q(t,\epsilon)$ is denoted by
\[
\delta q(t):= \left.\frac{d}{d\epsilon}\right|_{\epsilon =0} q(t,\epsilon).
\]
From the fixed endpoint conditions, we have $\delta q(t_1)=\delta q(t_2)=0$.

The Hamilton principle \eqref{HP_epsilon} is usually written shortly as
\begin{equation}\label{HP}
\delta \int_{t_1}^{t_2} L(q,  \dot{q}){\rm d}t=0,
\end{equation}
for arbitrary infinitesimal variations $\delta q$ with $\delta q(t_1)=\delta q(t_2)=0$. Throughout this paper, we shall always use this short notation for the variational principles and also simply refer to $\delta q$ as variations.
\medskip

A direct application of \eqref{HP_epsilon} gives, in local coordinates $q=(q^{1},...,q^{n})$,
\begin{equation}\label{computations_HP}
\begin{aligned}
\delta \int_{t_1}^{t_2} L(q, \dot{q}){\rm d}t&= \int_{t_1}^{t_2}\left[ \frac{\partial L}{\partial q^i}  \delta q^i + \frac{\partial L}{\partial \dot{q}^i}  \delta \dot{q}^i\right]{\rm d}t\\
&=\int_{t_1}^{t_2}\left[ \frac{\partial L}{\partial q^i}  -\frac{d}{dt} \frac{\partial L}{\partial \dot{q}^i}\right]   \delta q^i \,{\rm d}t + \left[ \frac{\partial L}{\partial \dot q^i}  \delta q^i\right]_{t_1}^{t_2},
\end{aligned}
\end{equation}
where we employ Einstein's summation convention.
Since $\delta q$ is arbitrary and since the boundary term vanishes because of the fixed endpoint  conditions, we get from \eqref{computations_HP} the \textit{Euler-Lagrange equations}
\begin{equation}\label{EL}
\frac{d}{dt}\frac{\partial L}{\partial \dot{q}^i}- \frac{\partial L}{\partial q^i}=0,\;\; i=1,...,n.
\end{equation}
We recall that $L$ is called \textit{regular}, when the Legendre transform $\mathbb{F}L:TQ \rightarrow T^*Q$, locally given by $(q^i, v^i) \mapsto (q^i,\frac{\partial L}{\partial v^i})$, is a local diffeomorphism, where $T^*Q$ denotes the cotangent bundle\footnote{The cotangent bundle of a manifold $Q$ is the manifold $T^*Q=\cup_{q\in Q}T^*_qQ$, where $T^*_qQ$ is the \textit{cotangent space} at each $q$ given as the dual space to $T_qQ$. The elements in $T^*_qQ$ are covectors, denoted by $(q,p)$.}, or {\it momentum phase space}, of $Q$. When $L$ is regular, the Euler-Lagrange equations \eqref{EL} yield a second-order differential equation for the curve $ q(t)$.

\medskip

The energy of a mechanical system with Lagrangian $L$ is defined on $TQ$ by
\begin{equation}\label{def_E_tot}
E(q, v)= \left<\frac{\partial L}{\partial v}, v\right>- L(q,v),
\end{equation}
where $\left<,\right>$ denotes a dual paring between the elements in $T^{\ast}_{q}Q$ and $T_{q}Q$.
It is easy to check that $E$ is conserved along the solutions of the Euler-Lagrange equations \eqref{EL}, namely,
\[
\frac{d}{dt}E(q, \dot{q})=\left(\frac{d}{dt}\frac{\partial L}{\partial \dot{q}^i}- \frac{\partial L}{\partial q^i}\right)  \dot{q}^i=0.
\]

Let us assume that the mechanical system is subject to an external force, given by a map $F^{\rm ext} :TQ \rightarrow T^* Q$ assumed to be fiber preserving, i.e. $F^{\rm ext}(q,v)\in T^*_qQ$ for all $(q,v)\in T_qQ$. The extension of \eqref{HP} to forced mechanical systems is given by
\begin{equation}\label{HP_force}
\delta \int_{t_1}^{t_2} L(q,  \dot{q}){\rm d}t+\int_{t_1}^{t_2} \left<F^{\rm ext}(q, \dot q), \delta q\right>\, {\rm d} t=0,
\end{equation}
for arbitrary variations $\delta q$ with $\delta q(t_1)=\delta q(t_2)=0$. The second term in \eqref{HP_force} is the time integral of the virtual work $ \left<F^{\rm ext}(q, \dot q), \delta q\right>$ done by the force field $F^{\rm ext}: TQ \to T^{\ast}Q$ with a virtual displacement $\delta q$ in $TQ$. The principle \eqref{HP_force} gives the \textit{forced Euler-Lagrange equations}
\begin{equation}\label{EL_force}
\frac{d}{dt}\frac{\partial L}{\partial \dot{q}^i}- \frac{\partial L}{\partial q^i}=F^{\rm ext}_i.
\end{equation}

\paragraph{Systems with nonholonomic constraints.} Hamilton's principle recalled above is only valid for \textit{holonomic} systems, i.e., systems without constraints or whose constraints are given by functions of the coordinates only, not the velocities. In geometric terms, such constraints are obtained by the specification of a submanifold $N$ of the configuration manifold $Q$. In this case, the equations of motion are still given by Hamilton's principle for the Lagrangian $L$ restricted to the tangent bundle $TN$ of the submanifold $N \subset Q$.

When the constraints cannot be reduced to relations between the coordinates only, they are called \textit{nonholonomic}. Here, we restrict the discussion to nonholonomic constraints that are linear in velocity. Such constraints are locally given in the form
\begin{equation}\label{linear_constraints}
\omega_i^\alpha(q) \dot q^i=0,\;\;\alpha=1,...,k <n,
\end{equation}
where $\omega_i^\alpha$ are functions of local coordinates $q=(q^{1},...,q^{n})$ on $Q$. Intrinsically the functions $\omega^\alpha_i$ are the components of $k$ independent one-forms $\omega^\alpha$ on $Q$, i.e., $\omega^\alpha = \omega_i^\alpha dq^i$, for $\alpha=1,...,k$.
Typical examples of linear nonholonomic constraints are those imposed on the motion of rolling bodies, namely that the velocities of the points in contact should be identical.

For systems with nonholonomic constraints \eqref{linear_constraints},  the corresponding equations of motion can be derived from a modification of the Hamilton principle called the \textit{Lagrange-d'Alembert principle}, which is given by
\begin{equation}\label{LdA_principle}
\delta \int_{t_1}^{t_2} L(q, \dot{q}){\rm d}t=0,
\end{equation}
for variations $\delta q$ subject to the condition
\begin{equation}\label{VC_mechanics}
\omega_i^\alpha(q) \delta q^i=0,\;\;\alpha=1,...,k<n,
\end{equation}
together with the fixed endpoint conditions $\delta q(t_1)=\delta q(t_2)=0$.
Note the occurrence of two constraints with distinct roles. First there is the constraint \eqref{linear_constraints} on the solution curve called the \textit{kinematic constraint}. Second, there is the constraint \eqref{VC_mechanics} on the variations used in the principle, referred to as the \textit{variational constraint}. Later, we show that this distinction becomes more noticeable in nonequilibrium thermodynamics.

A direct application of \eqref{LdA_principle}--\eqref{VC_mechanics} yields the \textit{Lagrange-d'Alembert equations}
\begin{equation}\label{LdA}
\frac{d}{dt}\frac{\partial L}{\partial \dot{q}^i}- \frac{\partial L}{\partial q^i}=\lambda_\alpha \omega^\alpha_i.
\end{equation}
These equations, together with the constraints equations \eqref{linear_constraints}, form a complete set of equations for the unknown curves $q^i(t)$ and $\lambda_\alpha(t)$.

For more information on nonholonomic mechanics, the reader can consult \cite{NeFu1969}, \cite{ArKoNe1988}, or \cite{Bl2003}. Note that the Lagrange-d'Alembert principle \eqref{LdA_principle} is \textit{not} a critical curve condition for the action integral \textit{restricted to the space of curve satisfying the constraints}. Such a principle, which imposes the constraint via a Lagrange multiplier, gives equations that are in general not equivalent to the Lagrange-d'Alembert equations \eqref{LdA}, see, e.g.,  \cite{LeMu1995}, \cite{Bl2003}. Such equations are sometimes referred to as the vakonomic equations.

\subsection{Continuum mechanics}\label{CLM}

Hamilton's principle admits a natural extension to continuum systems such as fluid and elasticity. For such systems the configuration manifold $Q$ is typically a manifold of maps. We shall restrict here to fluid mechanics in a fixed domain $\mathcal{D}\subset \mathbb{R}^3$, assumed to be bounded with smooth boundary $\partial \mathcal{D}$. Hamilton's principle for fluid mechanics in the Lagrangian description has been discussed at least since the works of \cite{He1955}, for an incompressible fluid and \cite{Se1959} and \cite{Ec1960} for compressible flows, see also \cite{TrTo1960} for further references on these early developments. Hamilton's principle has since then been an important modelling tool in continuum mechanics.

\paragraph{Configuration manifolds.} For fluid mechanics in a fixed domain, and before the occurrence of any shocks, the configuration space can be taken as the manifold $Q=\operatorname{Diff}(\mathcal{D})$\footnote{In this paper, we do not describe the functional analytic setting needed to rigorously work in the framework of infinite dimensional manifolds. For example, one can assume that the diffeomorphisms are of some given Sobolev class, regular enough (at least of class $C^1$), so that $\operatorname{Diff}(\mathcal{D})$ is a
smooth infinite dimensional manifold and a topological group with smooth right translation.} of diffeomorphisms of $\mathcal{D}$. The tangent bundle to $\operatorname{Diff}(\mathcal{D})$ is formally given by the set of vector fields on $\mathcal{D}$ covering a diffeomorphism $\varphi$ and tangent to the boundary, i.e., for each $\varphi\in\operatorname{Diff}(\mathcal{D})$, we have
\[
T_\varphi \operatorname{Diff}(\mathcal{D})= \{\mathbf{V}:\mathcal{D}\rightarrow T\mathcal{D}\mid \mathbf{V}(X)\in T_{\varphi(X)}\mathcal{D},\;\forall\; X\in \mathcal{D} ,\;\mathbf{V}(X)\in T_{\varphi(X)}\partial\mathcal{D},\;\forall\; X\in\partial\mathcal{D} \}.
\]

The motion of the fluid is fully described by a curve $\varphi_t\in \operatorname{Diff}(\mathcal{D})$ giving the position $x=\varphi_t(X)$ at time $t$ of a fluid particle with label $X\in \mathcal{D}$. The vector field $\mathbf{V}_t \in T_{\varphi_t}\operatorname{Diff}(\mathcal{D})$ defined by $\mathbf{V}_t(X)= \frac{d}{dt}\varphi_t(X)$ is the material velocity of the fluid.
In local coordinates, we write $x^a=\varphi_t^a(X^A)$ and $\mathbf{V}^a_t(X^A)=\frac{d}{dt}\varphi^a_t(X^A)$.

\paragraph{Hamilton's principle.} Given a Lagrangian $L:T Q \rightarrow \mathbb{R}$ defined on the tangent bundle of the infinite dimensional manifold $Q= \operatorname{Diff}(\mathcal{D})$, Hamilton's principle takes formally the same form as \eqref{HP}, namely 
\begin{equation}\label{HP_continuum}
\delta \int_{t_1}^{t_2} L(\varphi, \dot{\varphi}){\rm d}t=0,
\end{equation}
for variations $\delta\varphi$ such that $\delta\varphi_{t_1}=\delta \varphi_{t_2}=0$. 

Let us consider a Lagrangian of the general form
\[
L(\varphi, \dot{\varphi})=\int_\mathcal{D}  \mathscr{L}\left(\varphi(X), \dot{\varphi}(X), \nabla\varphi(X)\right){\rm d}^3X,
\]
with $\mathscr{L}$ the Lagrangian density and $\nabla\varphi$ the Jacobian matrix of $\varphi$, known as the deformation gradient in continuum mechanics. The variation of the integral yields
\begin{align*}
\delta\int_{t_1}^{t_2}L(\varphi, \dot{\varphi}){\rm d}t&= \int_{t_1}^{t_2}\!\!\int_\mathcal{D}\left[\frac{\partial\mathscr{L}}{\partial\varphi^a}\delta\varphi^a+\frac{\partial\mathscr{L}}{\partial\dot\varphi^a}\delta\dot\varphi^a+ \frac{\partial\mathscr{L}}{\partial\varphi^a_{,A}}\delta\varphi_{,A}^a \right]{\rm d}^3X{\rm d}t\\
&=\int_{t_1}^{t_2}\!\!\int_\mathcal{D}\left[ \frac{\partial\mathscr{L}}{\partial\varphi^a}\delta\varphi^a-\frac{\partial}{\partial t} \frac{\partial\mathscr{L}}{\partial\dot\varphi^a} - \frac{\partial }{\partial A}\frac{\partial\mathscr{L}}{\partial\varphi^a_{,A}}\right]\delta\varphi^a {\rm d}^3X{\rm d}t \\
&\qquad\qquad+ \int_{\mathcal{D}}\left[\frac{\partial\mathscr{L}}{\partial\dot\varphi^a}\delta \varphi^a \right]_{t_1}^{t_2} {\rm d}^3X + \int_{t_1}^{t_2}\!\!\int_{\partial \mathcal{D}} \frac{\partial\mathscr{L}}{\partial\varphi^a_{,A}} \mathbf{N}_A\delta \varphi^a {\rm d}{\mathcal S}{\rm d}t,
\end{align*}
where $\mathbf{N}$ is the outward pointing unit normal vector field to the boundary $\partial\mathcal{D}$ and ${\rm d}\mathcal S$ denotes the area element on the surface $\partial \mathcal{D}$.
Hamilton's principle thus yields the Euler-Lagrange equations and the boundary condition
\begin{equation}\label{EL_continuum}
\frac{\partial}{\partial t}\frac{\partial \mathscr{L}}{\partial \dot\varphi} + \operatorname{DIV} \frac{\partial \mathscr{L}}{\partial \nabla\varphi}= \frac{\partial \mathscr{L}}{\partial \varphi}\qquad\text{and}\qquad  \left.\frac{\partial\mathscr{L}}{\partial\nabla\varphi} \cdot \mathbf{N}\right|_{T\partial \mathcal{D}}=0\;\;\text{on $\partial\mathcal{D}$},
\end{equation}
where the divergence operator is defined as $\big( \operatorname{DIV} \frac{\partial \mathscr{L}}{\partial \nabla\varphi}\big)_a=\frac{\partial }{\partial A}\frac{\partial\mathscr{L}}{\partial\varphi^a_{,A}}$. The tensor field
\begin{equation}\label{PK}
\mathbf{P}:= - \frac{\partial \mathscr{L}}{\partial \nabla\varphi},\quad\text{i.e.}\quad \mathbf{P}^A_a= -  \frac{\partial\mathscr{L}}{\partial\varphi^a_{,A}} 
\end{equation}
is called the \textit{first Piola-Kirchhoff stress tensor}, see, e.g., \cite{MaHu1983}.

\paragraph{The Lagrangian of the compressible fluid.} For a compressible fluid, the Lagrangian has the standard form
\begin{equation}\label{L_fluid}
\begin{aligned}
L(\varphi, \dot \varphi)&=K(\varphi, \dot{\varphi})- U(\varphi)\\
&=\int_\mathcal{D} \left[ \frac{1}{2} \varrho_{\rm ref}(X) |\dot\varphi(X)|^2 - \mathscr{E}\big(\varrho_{\rm ref}(X),S_{\rm ref}(X),\nabla\varphi(X)\big)\right]{\rm d}^3X,
\end{aligned}
\end{equation}
with $\varrho_{\rm ref}(X)$ and $S_{\rm ref}(X)$ the mass density and entropy density in the reference configuration.
The two terms in \eqref{L_fluid} are, respectively, the total kinetic energy of the fluid and minus the total internal energy of the fluid. The function $\mathscr{E}$ is a general expression for the internal energy density written in terms of $\varrho_{\rm ref}(X)$, $S_{\rm ref}(X)$, and the deformation gradient $\nabla \varphi(X)$. For fluids $\mathscr{E}$ depends on the deformation gradient only through the Jacobian of $\varphi$ denoted $J_{\varphi}$. This fact is compatible with the material covariance property of $\mathscr{E}$, written as
\begin{equation}\label{Material_covariance}
\mathscr{E}\big(\psi^*\varrho_{\rm ref},\psi^*S_{\rm ref},\nabla(\varphi\circ \psi)\big)= \psi^*\big[\mathscr{E}\big(\varrho_{\rm ref},S_{\rm ref},\nabla \varphi \big)\big],\quad\text{for all $\psi\in \operatorname{Diff}(\mathcal{D})$},
\end{equation}
where the  pull-back notation is defined as
\begin{equation}\label{pull_back}
\varphi^* f= (f\circ\varphi)J_\varphi
\end{equation}
for some function $f$ defined on $\mathcal{D}$. From \eqref{Material_covariance} we deduce the existence of a function $\epsilon$ such that
\begin{equation}\label{mat_cov}
\mathscr{E}\big( \varrho_{\rm ref}, S_{\rm ref},\nabla \varphi\big)=\varphi^*\big[\epsilon (\rho,s) \big],\quad \text{for}\quad \rho=\varphi_*\varrho_{\rm ref},\quad s=\varphi_*S_{\rm ref},
\end{equation}
see \cite{MaHu1983}, \cite{GBMaRa2012}. The function $\epsilon=\epsilon(\rho,s)$ is the internal energy density in the spatial description, expressed in terms of the mass density $\rho$ and entropy density $s$ in the spatial description.

For the Lagrangian \eqref{L_fluid} and with the assumption \eqref{Material_covariance}, the first Piola-Kirchhoff stress tensor \eqref{PK} and its divergence are computed as
\begin{equation}\label{Piola_fluid}
\mathbf{P}^A_a=\frac{\partial \mathscr{E}}{\partial \varphi^a_{,A}}= - p J_\varphi (\varphi^{-1})^A_{,a}\quad\text{and}\quad \operatorname{DIV}\mathbf{P}= (\nabla p\circ\varphi) J_\varphi,
\end{equation}
where $p= \frac{\partial\epsilon}{\partial\rho} \rho+\frac{\partial\epsilon}{\partial s} s - \epsilon$ is the pressure. 
Note that for all $\delta\varphi^a$ parallel to the boundary, we have $\mathbf{P}^A_a\mathbf{N}_A\delta\varphi^a= - p J_\varphi (\varphi^{-1})^A_{,a}\mathbf{N}_A\delta\varphi^a=0$, since $(\varphi^{-1})^A_{,a}\delta\varphi^a$ is parallel to the boundary. Hence the boundary condition in \eqref{EL_continuum} is always satisfied. From \eqref{Piola_fluid} the Euler-Lagrange equations \eqref{EL_continuum} become
\begin{equation}\label{EL_material_fluid}
\varrho_{\rm ref}\ddot\varphi = (\nabla p\circ\varphi) J_\varphi.
\end{equation}
Equations \eqref{EL_material_fluid} are the equations of motion for a compressible fluid in the \textit{material (or Lagrangian) description}, which directly follows from the Hamilton principle \eqref{HP_continuum} applied to the Lagrangian \eqref{L_fluid}. It is however highly desirable to have a variational formulation that directly produces the equations of motion in the standard spatial (or Eulerian) description. This is recalled below in \S\ref{sec_reduction} by using Lagrangian reduction by symmetry.



\subsection{Lagrangian reduction by symmetry}\label{sec_reduction}

When a symmetry is available in a mechanical system, it is often possible to exploit it in order to reduce the dimension of the system and thereby facilitating its study. This process, called \textit{reduction by symmetry}, is today well understood both on the Lagrangian and Hamiltonian sides, see \cite{MaRa1999} for an introduction and references.

\medskip

While on the Hamiltonian side, this process is based on the  reduction of symplectic or Poisson structures, on the Lagrangian side it is usually based on the reduction of variational principles, see \cite{MaSc1993a,MaSc1993b}, \cite{CeMaRa2001}. Consider a mechanical system with configuration manifold $Q$ and Lagrangian $L:TQ\rightarrow\mathbb{R}$ and consider also the action of a Lie group $G$ on $Q$, denoted here simply as $q\mapsto g\cdot q$, for $g\in G$, $q\in Q$. This action naturally induces an action on the tangent bundle $TQ$, denoted here simply as $(q,v) \mapsto (g\cdot q, g\cdot v)$, called the \textit{tangent lifted action}. We say that the action is a symmetry for the mechanical system if the Lagrangian $L$ is invariant under this tangent lifted action. In this case, $L$ induces a \textit{symmetry reduced Lagrangian} $\ell: (TQ)/G\rightarrow\mathbb{R}$ defined on the quotient space $(TQ)/G$ of the tangent bundle with respect to the action. The goal of the Lagrangian reduction process is to derive the equations of motion directly on the reduced space $(TQ)/G$. Under standard hypotheses on the action, this quotient space is a manifold and one obtains the \textit{reduced Euler-Lagrange equations} by computing the \textit{reduced variational principle} for the action integral $\int_{t_1}^{t_2}\ell\,{\rm d}t$ induced by Hamilton's principle \eqref{HP} for the action integral $\int_{t_1}^{t_2}L\,{\rm dt}$. The main difference between the reduced variational principle and Hamilton's principle is the occurrence of constraints on the variations to be considered when computing the critical curves for $\int_{t_1}^{t_2}\ell\,{\rm d}t$. These constraints are uniquely associated to the reduced character of the variational principle and are not due to physical constraints as in \eqref{VC_mechanics} earlier.


\medskip

We now quickly recall the application of Lagrangian reduction for the treatment of fluid mechanics in a fixed domain, see \S\ref{CLM}, by following the Euler-Poincar\'e reduction approach in \cite{HoMaRa1998}. In this case the Lagrangian reduction process encodes the passing from the material (or Lagrangian) description to the spatial (or Eulerian) description.

As we recalled above, in the material description the motion of the fluid is described by a curve of diffeomorphisms $\varphi_t$ in the configuration manifold $Q=\operatorname{Diff}(\mathcal{D})$ and the evolution equations \eqref{EL_material_fluid} for $\varphi_t$ follow from the standard Hamilton principle.

In the spatial description, the dynamics is described by the Eulerian velocity $\mathbf{v}(t,x)$, the mass density $\rho(t,x)$ and the entropy density $s(t,x)$ defined in terms of $\varphi_t$ as
\begin{equation}\label{def_Eulerian}
\mathbf{v}_t=\dot\varphi_t\circ\varphi_t^{-1},\quad \rho_t=( \varphi_t)_*\varrho_{\rm ref},\quad s_t=(\varphi_t)_*S_{\rm ref}.
\end{equation}
Using these relations and \eqref{mat_cov}, the Lagrangian \eqref{L_fluid} in material description induces the following expression in the spatial description
\[
\ell(\mathbf{v}, \rho,s)= \int_\mathcal{D}\left[\frac{1}{2}\rho |\mathbf{v}|^2 - \varepsilon (\rho,s) \right]{\rm d}^3x.
\]
The symmetry group underlying the Lagrangian reduction process is the subgroup
\[
\operatorname{Diff}(\mathcal{D})_{\varrho_{\rm red}, S_{\rm ref}}\subset\operatorname{Diff}(\mathcal{D})
\]
of diffeomorphisms that preserve both the mass density $\varrho_{\rm ref}$ and entropy density $S_{\rm ref}$ in the reference configuration. So, we have $Q= \operatorname{Diff}(\mathcal{D})$ and $G=\operatorname{Diff}(\mathcal{D})_{\varrho_{\rm red}, S_{\rm ref}}$ in the general Lagrangian reduction setting described above.

From the relations \eqref{def_Eulerian} we obtain that the variations $\delta\varphi$ used in Hamilton's principle \eqref{HP_continuum} induce the variations
\begin{equation}\label{variations_Eulerian}
\delta\mathbf{v}=\partial_t\boldsymbol{\zeta} + \mathbf{v}\cdot\nabla \boldsymbol{\zeta} - \boldsymbol{\zeta} \cdot \nabla\mathbf{v},\quad \delta\rho=-\operatorname{div}(\rho \boldsymbol{\zeta} ), \quad \delta s = - \operatorname{div}(s\boldsymbol{\zeta} ),
\end{equation}
where $\boldsymbol{\zeta}=\delta\varphi\circ\varphi^{-1}$ is an arbitrary time dependent vector field parallel to $\partial\mathcal{D}$. From Lagrangian reduction theory the Hamilton principle \eqref{HP_continuum} induces, in the Eulerian description, the (reduced) variational principle
\begin{equation}\label{EP}
\delta\int_{t_1}^{t_2}\ell(\mathbf{v}, \rho,s)\,{\rm d}t=0,
\end{equation}
for variations $\delta\mathbf{v}$, $\delta\rho$, $\delta s$ constrained by the relations \eqref{variations_Eulerian} with $\boldsymbol{\zeta}(t_1)=\boldsymbol{\zeta}(t_2)=0$. This principle yields the compressible fluid equations $\rho(\partial_t\mathbf{v}+ \mathbf{v}\cdot\nabla\mathbf{v})=- \nabla p$ in the Eulerian description, while the continuity equations $\partial_t\rho+ \operatorname{div}(\rho\mathbf{v})=0$ and $\partial_t s+ \operatorname{div}( s\mathbf{v})=0$ follow from the definition of $\rho$ and $s$ in \eqref{def_Eulerian}, see \cite{HoMaRa1998}. We refer to \cite{GBMaRa2012} for extension of this Lagrangian reduction approach to the case of fluids with a free boundary. 

The variational formulation \eqref{variations_Eulerian}--\eqref{EP} are extended in \S\ref{sec_continuum} to include irreversible processes and are illustrated using the Navier-Stokes-Fourier system as an example.

\section{Variational formulation for discrete thermodynamic systems}\label{sec_finite}

In this section we present a variational formulation for the finite dimensional case of discrete thermodynamic systems that reduces to Hamilton's variational principle \eqref{HP} in absence of irreversible processes. The form of this variational formulation is similar to that of nonholonomic mechanics recalled earlier, see \eqref{linear_constraints}--\eqref{VC_mechanics}, in the sense that the critical curve condition is subject to two constraints: a \textit{kinematic constraints} on the solution curve and a \textit{variational constraint} on the variations to be considered when computing the criticality condition. A major difference however, with the Lagrange-d'Alembert principle recalled above, is that the constraints are \textit{nonlinear} in velocities. This formulation is extended to continuum systems in \S\ref{sec_continuum}.

\medskip

Before presenting the variational formulation, we recall below the two laws of thermodynamics as formulated in \cite{StSc1974}.

\paragraph{The two laws of thermodynamics.}
Let us denote by $ \boldsymbol{\Sigma}  $ a physical system and by $ \boldsymbol{\Sigma} ^{\rm ext}$ its exterior. The state of the system is defined by a set of mechanical variables and a set of thermal variables. State functions are functions of these variables. Stueckelberg's formulation of the two laws is given as follows.

\medskip

\noindent {\bf First law:} For every system $ \boldsymbol{\Sigma} $, there exists an extensive scalar state function $E$, called \textit{\textbf{energy}}, which satisfies
\[
\frac{d}{dt} E(t) = P^{\rm ext}_W(t)+P^{\rm ext}_H(t)+P^{\rm ext}_M(t),
\]
where $ P^{\rm ext}_W$ is the {\it power associated to the work\footnote{Here {\it work} includes not only {\it mechanical work} by the action of forces but also other physical work such as the one by the action of electric voltages, etc.} done on the system}, $P^{\rm ext}_H$ is the {\it power associated to the transfer of heat into the system}, and $P^{\rm ext}_M$ is the {\it power associated to the transfer of matter into the system}.\footnote{As we recall below, to a transfer of matter into the system is associated a transfer of  work and heat. By convention, $ P^{\rm ext}_W$ and $P^{\rm ext}_H$ denote uniquely the power associated to a transfer of work and heat into the system that is \textit{not associated to a transfer of matter}. The power associated to a transfer of heat or work due to a transfer of matter is included in $P^{\rm ext}_M$.}
\medskip

Given a {\it thermodynamic system}, the following terminology is generally adopted:
\begin{itemize}
\item A system is said to be \textit{\textbf{closed}} if there is no exchange of matter, i.e.,  $P^{\rm ext}_M(t)=0$. When $P^{\rm ext}_M(t) \ne 0$ the system is said to be \textit{\textbf{open}}.
\item 
A system is said to be \textit{\textbf{adiabatically closed}} if it is closed and there is no heat exchanges, i.e., $P^{\rm ext}_M(t)=P^{\rm ext}_H(t)=0$. 
\item 
A system is said to be \textit{\textbf{isolated}} if it is adiabatically closed and there is no mechanical power exchange, i.e., $P^{\rm ext}_M(t)=P^{\rm ext}_H(t)=P^{\rm ext}_W(t)=0$.
\end{itemize}

From the first law, it follows that the {\it energy of an isolated system is constant}.
\medskip

\noindent {\bf Second law:} For every system $ \boldsymbol{\Sigma} $, there exists an extensive scalar state function $S$, called \textit{\textbf{entropy}}, which obeys the following two conditions
\begin{itemize}
\item[(a)]  Evolution part:\\
If the system is adiabatically closed, the entropy $S$ is a non-decreasing function with respect to time, i.e., 
\[
\frac{d}{dt} S(t)=I(t)\geq 0,
\]
where $I(t)$ is the {\it entropy production rate} of the system accounting for the irreversibility of internal processes.
\item[(b)] Equilibrium part:\\
If the system is isolated, as time tends to infinity the entropy tends towards a finite local maximum of the function $S$ over all the thermodynamic states $ \rho $ compatible with the system, i.e., 
\[
\lim_{t \rightarrow +\infty}S(t)= \max_{ \rho \; \text{compatible}}S[\rho ].
\]
\end{itemize}

By definition, the evolution of an isolated system is said to be {\it reversible} if $I(t)=0$, namely, the entropy is constant. In general, the evolution of a system $ \boldsymbol{\Sigma} $ is said to be {\it reversible}, if the evolution of the total isolated system with which $ \boldsymbol{\Sigma} $ interacts is reversible.

\medskip

Based on this formulation of the two laws,  \cite{StSc1974} developed a systematic approach for the derivation of the equations of motion for thermodynamic systems, especially well suited for the understanding of nonequilibrium thermodynamics as an extension of classical mechanics. We refer, for instance, to \cite{Gr1999}, \cite{FeGr2010}, \cite{GrBr2011} for the applications of Stueckelberg's approach to the derivation of equations of motion for thermodynamical systems.

\medskip

We shall present our approach by considering systems with gradually increasing level of complexity. First we treat \textit{adiabatically closed} systems, which have only one entropy variable or, equivalently, one temperature. Such systems, called \textit{simple systems}, may involve the irreversible processes of mechanical friction and internal matter transfer. Then we treat a more general class of finite dimensional adiabatically closed thermodynamic systems with several entropy variables, which may also involve the irreversible process of heat conduction. We then consider \textit{open} finite dimensional thermodynamic systems, which can exchange heat and matter with the exterior. Finally, we explain how chemical reactions can be included in the variational formulation.

\subsection{Adiabatically closed simple thermodynamic systems}\label{subsec_simple}

We present below the definition of finite dimensional and simple systems following \cite{StSc1974}.
A \textit{finite dimensional thermodynamic system} $ \boldsymbol{\Sigma} $ is a collection $ \boldsymbol{\Sigma} = \cup_{A=1}^P\boldsymbol{\Sigma} _A$ of a finite number of interacting simple thermodynamic systems $ \boldsymbol{\Sigma} _A $. By definition, a \textit{simple thermodynamic system} is a macroscopic system for which one (scalar) thermal variable and a finite set of non thermal variables are sufficient to describe entirely the state of the system. From the second law of thermodynamics, we can always choose  the entropy $S$ as a thermal variable. A typical example of such a simple system is the one-cylinder problem. We refer to \cite{Gr1999} for a systematic treatment of this system via Stueckelberg's approach. 

\paragraph{(A) Variational formulation for mechanical systems with friction.} We consider here a simple system in which the system can be described only by a single entropy as a thermodynamic variable, beside mechanical variables. As in \S\ref{sec_LM} above, let $Q$ be the configuration manifold associated to the mechanical variables of the simple system. The Lagrangian of the simple thermodynamic system is thus a function
\[
L: TQ \times \mathbb{R}  \rightarrow \mathbb{R} , \quad (q, v, S) \mapsto L(q, v, S),
\]
where $S \in\mathbb{R}$ is the entropy. We assume that the system involves external and friction forces given by fiber preserving maps $F^{\rm ext}, F^{\rm fr}:TQ\times \mathbb{R} \rightarrow T^* Q$, i.e., such that $F^{\rm fr}(q, v, S)\in T^*_qQ$, similarly for $F^{\rm ext}$. As stated in \cite{GBYo2017a}, the variational formulation for this simple system is given as follows:

\medskip
\begin{framed}
\noindent Find the curves $q(t)$, $S(t)$ which are critical for the \textit{variational condition}
\begin{equation}\label{LdA_thermo_simple} 
\delta \int_{t _1 }^{ t _2}L(q , \dot q , S){\rm d}t +\int_{t_1}^{t_2} \left<F^{\rm ext}(q, \dot q, S), \delta q \right>\,{\rm d}t =0,
\end{equation}
subject to the \textit{phenomenological constraint}
\begin{equation}\label{CK_simple} 
\frac{\partial L}{\partial S}(q, \dot q, S)\dot S  =  \left<F^{\rm fr}(q, \dot q, S), \dot q \right>,\qquad 
\end{equation}
and for variations subject to the  \textit{variational constraint}
\begin{equation}\label{CV_simple} 
\frac{\partial L}{\partial S}(q, \dot q, S)\delta S=  \left<F^{\rm fr}(q , \dot q , S), \delta q \right>,
\end{equation}
with $ \delta q(t_1)=\delta q(t_2)=0$.
\end{framed}

\medskip

Taking variations of the integral in \eqref{LdA_thermo_simple}, integrating by parts, and using $ \delta q(t_1)=\delta (t_2)=0$, it follows
\[
\int_{t_1}^{t_2}\left[ \left(\frac{\partial L}{\partial q^i}  -\frac{d}{dt} \frac{\partial L}{\partial \dot{q}^i}+F^{\rm ext}_i\right)   \delta q^i + \frac{\partial L}{\partial S} \delta S\right]\,{\rm d}t.
\]
From the variational constraint \eqref{CV_simple}, the last term in the integrand of the above equation can be replaced by $F^{\rm fr}_i\delta q^i$. Hence, using \eqref{CK_simple}, we get the following system of evolution equations for the curves $q(t)$ and $S(t)$:
\begin{equation}\label{simple_systems} 
\left\{
\begin{array}{l}
\displaystyle\vspace{0.2cm}\frac{d}{dt}\frac{\partial L}{\partial \dot q}- \frac{\partial L}{\partial q}=  F^{\rm fr}(q, \dot q, S)+F^{\rm ext}(q, \dot q, S),\\
\displaystyle\frac{\partial L}{\partial S}\dot S=  \left<F^{\rm fr}(q, \dot q, S), \dot q \right>.
\end{array} \right.
\end{equation} 
This variational formulation is a generalization of Hamilton's principle in Lagrangian mechanics in the sense that it can yield the irreversible processes in addition to the Lagrange-d'Alembert equations with external and friction forces. In this generalized variational formulation, the temperature is defined as minus the derivative of $L$ with respect to $S$, i.e., $T=-\frac{\partial L}{\partial S}$, which is assumed to be positive. When the Lagrangian has the standard form
\[
L(q,v, S)= K(q, v ) - U(q,S),
\]
where the kinetic energy $K$ is assumed to be independent on $S$ and $U(q,S)$ is the internal energy, then $T=-\frac{\partial L}{\partial S}=\frac{\partial U}{\partial S}$, recovers the standard definition of the temperature in thermodynamics.

\medskip
 
When the friction force vanishes, the entropy is constant from the second equation in \eqref{simple_systems}, and hence the system \eqref{simple_systems} reduces to the forced Euler-Lagrange equations  in classical mechanics, for a Lagrangian depending parametrically on a given constant entropy $S_0$.

\medskip

The total energy associated with the Lagrangian is still defined with the same expression as in \eqref{def_E_tot} except that now it depends on $S$, i.e., we define the total energy $E:TQ\times\mathbb{R}\rightarrow\mathbb{R}$ by
\begin{equation}\label{def_E_tot_thermo}
E(q,v,S)= \left<\frac{\partial L}{\partial v},  v\right>- L(q, v,S).
\end{equation}
Along the solution curve of \eqref{simple_systems}, we have
\[
\frac{d}{dt} E = \left(\frac{d}{dt}\frac{\partial L}{\partial \dot{q}^i}- \frac{\partial L}{\partial q^i}\right)  \dot{q}^i - \frac{\partial L}{\partial S}\dot S= F^{\rm ext}_i \dot q ^i= P^{\rm ext}_W,
\]
where $P_W^{\rm ext}$ is the power associated to the work done on the system. This is nothing but the statement of the first law for the thermodynamic system as in \eqref{simple_systems}.
\medskip

The rate of entropy production of the system is
\[
\dot S= -\frac{1}{T}\left<F^{\rm fr},\dot q\right>.
\]
The second law states that the internal entropy production is always positive, from which the friction force is dissipative, i.e., $\left<F^{\rm fr}(q, \dot q, S),\dot q \right>\leq 0$ for all $(q,\dot q,S)$. This suggests the phenomenological relation $F^{\rm fr}_{i}=-\lambda_{ij} \dot{q}^{j}$, where $\lambda_{ij}$, $i,j=1,...,n$ are functions of the state variables with the symmetric part of the matrix $\lambda_{ij}$ positive semi-definite, which are determined by experiments.

\begin{remark}[Phenomenological and variational constraints]\label{PC_VS_VC}{\rm 
The explicit expression of the constraint \eqref{CK_simple} involves phenomenological laws for the friction
force $F^{\rm fr}$, this is why we refer to it as a \textit{phenomenological constraint}. The associated constraint \eqref{CV_simple}  is called a \textit{variational
constraint} since it is a condition on the variations to be used in \eqref{LdA_thermo_simple}. Note that the constraint \eqref{CK_simple} is nonlinear and also that one passes from the variational constraint to the phenomenological constraint by formally replacing the time derivatives $ \dot q$, $\dot S$ by the variations $ \delta q$, $\delta S$:
\[
\frac{\partial L}{\partial S} \dot S  =  \left<F^{\rm fr}, \dot q \right> \;\;\leadsto\;\;\frac{\partial L}{\partial S} \delta S  =  \left<F^{\rm fr},  \delta q \right>.
\]
Such a systematic correspondence between the phenomenological and variational constraints will hold, in general, for our variational formulation of thermodynamics, as we present in detail below.}
\end{remark}

\begin{remark}{\rm In our macroscopic description, it is assumed that the macroscopically ``{\it slow\,}'' or collective motion of the system can be described by $q(t)$, while the time evolution of the entropy $S(t)$ is determined from the microscopically ``{\it fast\,}'' motions of molecules through statistical mechanics under the assumption of  \textit{local equilibrium}.
It follows from statistical mechanics that the internal energy $U(q,S)$, given as a potential energy at the macroscopic level, is essentially coming from the total kinetic energy associated with the microscopic motion of molecules, which is directly related to the temperature of the system.}
\end{remark}

\begin{figure}[h]
\begin{center}
\includegraphics[scale=0.77]{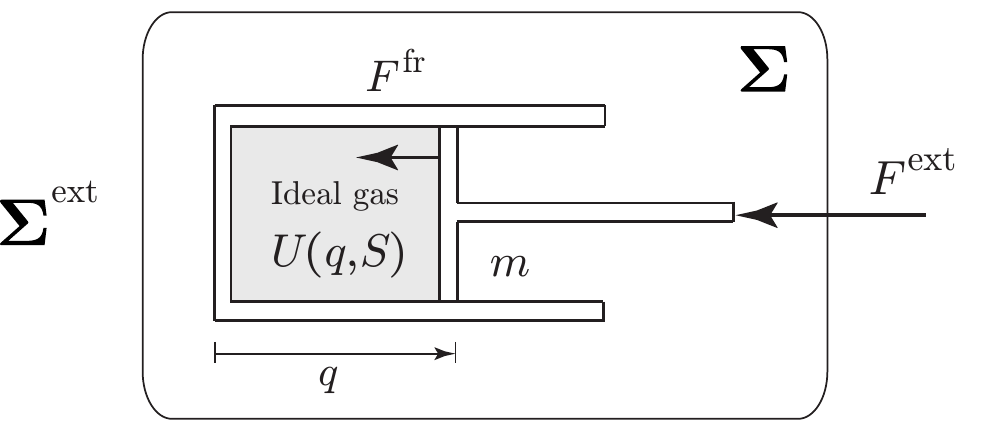}
\caption{One-cylinder.}
\label{one_cylinder}
\end{center}
\end{figure}

\medskip
\noindent\textbf{Example: piston.} Consider a gas confined by a piston in a cylinder as in Fig.\,\ref{one_cylinder}. This is an example of a simple adiabatically closed system, whose state can be characterized by $(q, v, S)$.
The Lagrangian is given by $L(q,v , S)=\frac{1}{2} m v^2 -U(q,S)$, where $m$ is the mass of the piston, $U(q,S):=\mathsf{U}(S,V=Aq,N_0)$, with $\mathsf{U}(S,V,N)$ the internal energy of the gas, $N_0$ is the constant number of moles, $V=\alpha q$ is the volume, and $\alpha$ is the constant area of the cylinder. Note that we have
\[
\frac{\partial U}{\partial S}(q,S)=T(q,S)\quad \text{and} \quad  \frac{\partial U}{\partial q}(q,S)=- p(q,S)\alpha,
\]
where $T$ is temperature and $p=- \frac{\partial \mathsf{U}}{\partial V}$ is the pressure.
The friction force reads $F^{\rm fr}(q, \dot q, S)=- \lambda (q, S) \dot q$, where $ \lambda (q, S)\geq 0$ is the phenomenological coefficient,  determined experimentally.

Following \eqref{LdA_thermo_simple}--\eqref{CV_simple}, the variational formulation is given by
\[
\delta \int_{t _1 }^{t _2 } \left[\frac{1}{2} m \dot q^2 -U(q,S)\right] {\rm d}t +\int_{ t _1 }^{ t _2 }  F^{\rm ext}(q, \dot q, S)  \delta q \,{\rm d}t=0,
\]
subject to the phenomenological constraint 
\[
\frac{\partial U}{\partial S}(q,S)\dot S= \lambda (q, S) \dot q^2.
\]
and for variations subject to the variational constraint
\[
\frac{\partial U}{\partial S}(q,S)\delta  S= \lambda (q, S) \dot q \delta q.
\]

From this principle, we get the equations of motion for the piston-cylinder system as
\[
m\ddot q =p(q,S)\alpha+F^{\rm ext}- \lambda (q, S)\dot q, \qquad T(q,S)\dot S= \lambda (q,  S)\dot q^2,
\]
consistently with the equations derived by \cite[\S4]{Gr1999}.  We can verify the energy balance, i.e., the first law, as $\frac{d}{dt} E=F^{\rm ext}\dot q$,
where $E=\frac{1}{2} m \dot q^2+ U$ is the total energy.
\medskip

\paragraph{(B) Variational formulation for systems with internal mass transfer.} We here extend the previous variational formulation to the finite dimensional case of discrete systems experiencing internal diffusion processes. Diffusion is particularly important in biology where many processes depend on the transport of chemical species through bodies. For instance, the setting that we develop is well appropriate for the description of diffusion across composite membranes, e.g., composed of different elements arranged in a series or parallel array, which occur frequently in living systems and have remarkable physical properties, see \cite{KeKa1963a,KeKa1963b,KeKa1963c,OsPeKa1973}.
\medskip

\begin{figure}[h]
\begin{center}
\includegraphics[scale=0.6]{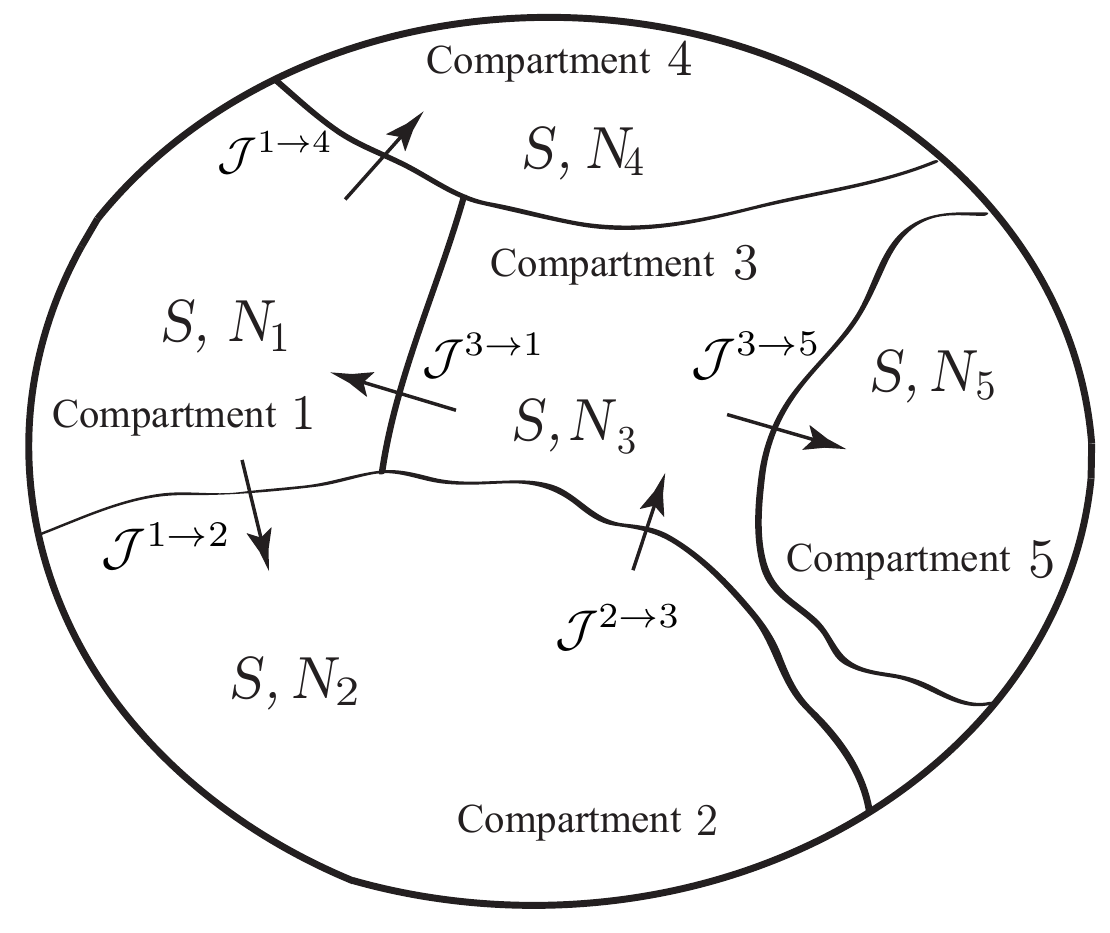}
\caption{Simple adiabatically closed system with a single chemical species, experiencing diffusion between several compartments.}
\label{ClosedMassTransSim}
\end{center}
\end{figure}

As illustrated in Fig.\,\ref{ClosedMassTransSim}, we consider a thermodynamic system consisting of $K$ compartments that can exchange matter by diffusion across walls (or membranes) of their common boundaries. We assume that the system has a single species and denote by $N _k $ the number of moles of the species in the $k$-th compartment, $k=1,...,K$. We assume that the thermodynamic system is simple; i.e., a uniform entropy $S$, the entropy of the system, is attributed to all the compartments.

For each compartment $k=1,...,K$, the mole balance equation is
\[
\frac{d}{dt} N _k = \sum_{\ell=1}^K \mathcal{J} ^{\ell \rightarrow k},
\]
where $\mathcal{J} ^{\ell \rightarrow k}=- \mathcal{J} ^{k \rightarrow \ell}$ is the molar flow rate from compartment $\ell$ to compartment $k$ due to diffusion of the species. We assume that the simple system also involves mechanical variables, friction and exterior forces $F^{\rm fr}$ and $F^{\rm ext}$, as in (A). The Lagrangian of the system is thus a function
\[
L: TQ\times\mathbb{R} \times \mathbb{R}^K  \rightarrow \mathbb{R} , \quad \left(q, v,S, N_1,...,N_K\right) \mapsto L\left(q, v,S, N_1,...,N_K\right).
\]

\paragraph{Thermodynamic displacements associated to matter exchange.} The variational formulation involves the new variables $W^k$, $k=1,...,K$, which are examples of \textit{thermodynamic displacements} and play a central role in our formulation. In general, we define the \textit{thermodynamic displacement associated to a irreversible process} as the primitive in time of the thermodynamic force (or affinity) of the process. This force (or affinity) thus becomes the rate of change of the thermodynamic displacement.
In the case of matter transfer, $\dot W^k$ corresponds to the chemical potential of $N_k$.

\medskip

The variational formulation for a simple system with internal diffusion process is stated as follows.

\begin{framed}
\noindent Find the curves $q(t)$, $S(t)$, $W^k(t)$, $N_k(t)$ which are critical for the \textit{variational condition}
\begin{equation}\label{VCond_simple_diffusion}
\delta \int_{t _1 }^{ t _2} \!\Big[ L\left(q, \dot q, S, N_1,...,N_K\right)+ \dot W^kN_k\Big] {\rm d}t + \int_{t_1}^{t_2} \left<\!F^{\rm ext }, \delta q\right>\,{\rm d}t=0, 
\end{equation}
subject to the \textit{phenomenological constraint}
\begin{equation}\label{PC_simple_diffusion}
\frac{\partial L}{\partial S}\dot S  =  \left<F^{\rm fr}, \dot q\right>   + \sum_{k,\ell=1}^{K}  \mathcal{J}^{\ell \rightarrow k} \dot W^k,
\end{equation}
and for variations subject to the \textit{variational constraint}
\begin{equation}\label{VC_simple_diffusion}
\frac{\partial L}{\partial S}\delta S  = \left< F^{\rm fr},\delta q \right> + \sum_{k,\ell=1}^{K}  \mathcal{J}^{\ell \rightarrow k} \delta W ^k,
\end{equation}
with $\delta q(t_1)=\delta q(t_2)=0$ and $ \delta W^k(t_1)=\delta W^k(t_2)=0$, $k=1,...,K$.
\end{framed}

\medskip

Taking variations of the integral in \eqref{VCond_simple_diffusion}, integrating by parts, and using $ \delta q(t_1)=\delta q(t_2)=0$ and $ \delta W^k(t_1)=\delta W^k(t_2)=0$, it follows
\[
\int_{t_1}^{t_2}\left[ \left(\frac{\partial L}{\partial q^i}  -\frac{d}{dt} \frac{\partial L}{\partial \dot{q}^i}+F^{\rm ext}_i\right)   \delta q^i + \frac{\partial L}{\partial S} \delta S + \left( \frac{\partial L}{\partial N_k} +\dot W^k\right)\delta N_k - \dot N_k \delta W^k\right]\,{\rm d}t.
\]
Then, using the variational constraint \eqref{VC_simple_diffusion}, we get the following conditions:
\begin{equation}
\begin{aligned}\label{OpenSystemEquation} 
\delta q^{i}:&\quad \frac{d}{dt}\frac{\partial L}{\partial \dot{q}^{i}}-\frac{\partial L}{\partial q^{i}}=F^{\rm fr}_{i}+F^{\rm ext }_{i}, \quad i=1,...,n,\\
\delta N_k:&\quad  \frac{d}{dt} W^{k}=-\frac{\partial L}{\partial N_k},\quad k=1,...,K,\\
\delta W^k:&\quad  \frac{d}{dt} N _k = \sum_{\ell=1}^K \mathcal{J} ^{\ell \rightarrow k}, \quad k=1,...,K.
\end{aligned}
\end{equation}
These conditions, combined with the phenomenological constraint \eqref{PC_simple_diffusion}, yield the system of evolution equations for the curves $q(t)$, $S(t)$, and $N^k(t)$:
\begin{equation}\label{simple_systems_matter} 
\left\{
\begin{array}{l}
\displaystyle\vspace{0.2cm}\frac{d}{dt}\frac{\partial L}{\partial \dot q}- \frac{\partial L}{\partial q}=  F^{\rm fr} +F^{\rm ext} ,\\
\displaystyle\vspace{0.2cm}\frac{d}{dt} N _k = \sum_{\ell=1}^K \mathcal{J} ^{\ell \rightarrow k}, \quad k=1,...,K,\\
\displaystyle\frac{\partial L}{\partial S}\dot S=  \left<F^{\rm fr},\dot q \right> - \sum_{k<\ell} \mathcal{J}^{\ell\rightarrow k}\left(\frac{\partial L}{\partial N_k} - \frac{\partial L}{\partial N_\ell}\right) .
\end{array} \right.
\end{equation} 


The total energy is defined as in \eqref{def_E_tot} and \eqref{def_E_tot_thermo} and depends here on the mechanical variables $(q,v)\in TQ$, the entropy $S$, and the number of moles $N_k$, $k=1,...,K$, i.e., we define $E:TQ\times\mathbb{R}\times\mathbb{R}^K\rightarrow\mathbb{R}$ as
\begin{equation}\label{def_E_tot_thermo_matter}
E\left(q, v,S, N_1,...,N_K\right)= \left<\frac{\partial L}{\partial v},  v\right>- L\left(q,v,S,N_1,...,N_K\right).
\end{equation}
On the solutions of \eqref{simple_systems_matter}, we have
\[
\frac{d}{dt} E= \left(\frac{d}{dt}\frac{\partial L}{\partial \dot{q}^i}- \frac{\partial L}{\partial q^i}\right)  \dot{q}^i - \frac{\partial L}{\partial S}\dot S- \frac{\partial L}{\partial N_k}\dot N_k= F^{\rm ext}_i\dot q ^i= P^{\rm ext}_W,
\]
where $P_W^{\rm ext}$ is the power associated to the work done on the system. This is the statement of the first law for the thermodynamic system \eqref{simple_systems_matter}.

\medskip

For a given Lagrangian $L$, the temperature and chemical potentials of each compartment are defined as
\[
T:=-\frac{\partial L}{\partial S}\qquad\text{and}\qquad \mu^k:=-\frac{\partial L}{\partial N_k},\;\; k=1,...,K.
\]
The last equation in \eqref{simple_systems_matter} yields the rate of entropy production of the system as
\[
\dot S=-\frac{1}{T} \left<F^{\rm fr}, \dot{q} \right>+ \frac{1}{T}\sum_{k<\ell}
\mathcal{J} ^{k \rightarrow \ell} (\mu^k -\mu^\ell),
\]
where the two terms correspond, respectively, to the rate of entropy production due to mechanical friction and to matter transfer.
The second law suggests the phenomenological relations
\[
F^{\rm fr}_{i}=-\lambda_{ij} \dot{q}^{j}\qquad\text{and}\qquad \mathcal{J} ^{k \rightarrow \ell}=G^{kl} (\mu ^k-\mu^\ell),
\]
where $\lambda_{ij}$, $i,j=1,...,n$ and $G^{k\ell}$, $k,\ell=1,...,K$ are functions of the state variables, with the symmetric part of the matrix $\lambda_{ij}$ positive semi-definite and with $G^{k\ell}\geq 0$, for all $k,\ell$.

\begin{figure}[h]
\begin{center}
\includegraphics[scale=0.55]{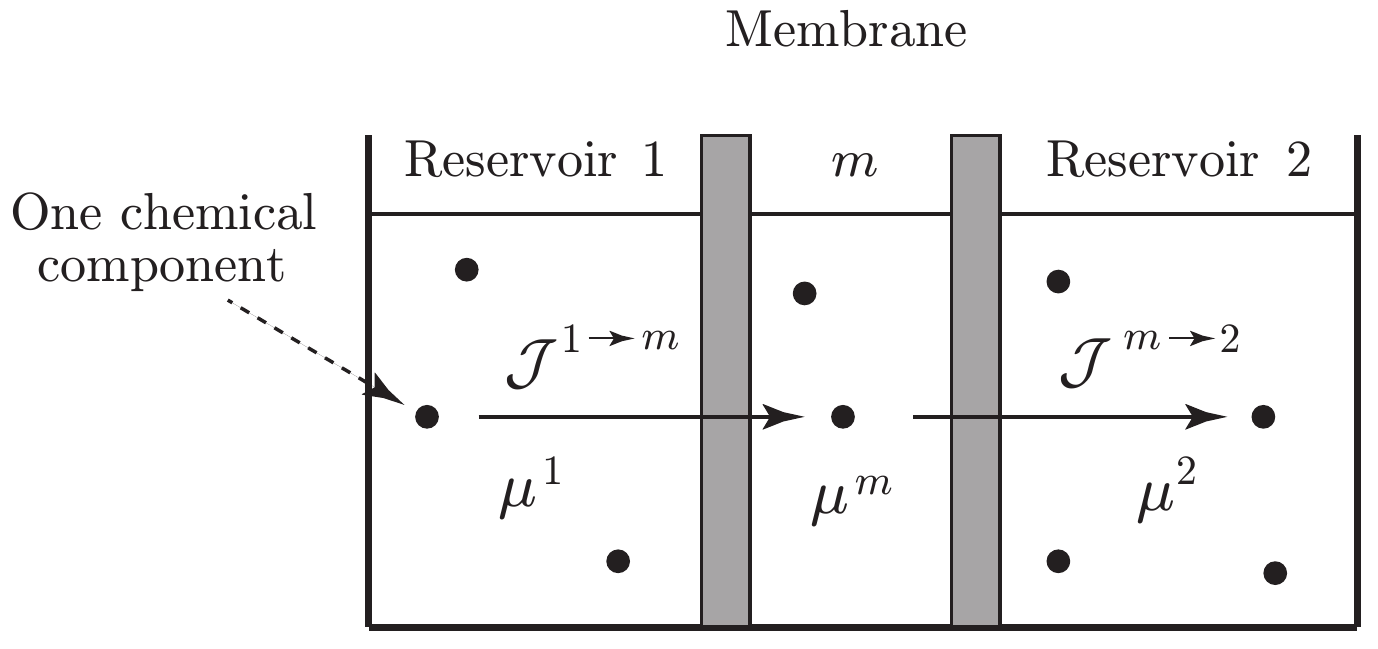}
\caption{Nonelectrolyte diffusion through a homogeneous membrane.}
\label{MatterTransport}
\end{center}
\end{figure}

\paragraph{Example: mass transfer associated to nonelectrolyte diffusion through a homogeneous membrane.}
We consider a {\it system with diffusion due to internal matter transfer} through a homogeneous membrane separating two reservoirs. We suppose that the system is simple (so it is described by a single entropy variable) and involves a single chemical species.  We assume that the membrane consists of three regions; namely, the central layer denotes the membrane capacitance in which energy is stored without dissipation, while the outer layers indicate transition regions in which dissipation occurs with no energy storage. We denote by $N_{m}$ the number of mole of this chemical species in the membrane and by $N_{1}$ and $N_{2}$ the numbers of mole in the reservoirs $1$ and $2$, as shown in Fig.\,\ref{MatterTransport}. Define the Lagrangian by $L(S, N_{1}, N_{2}, N_{m})=-U(S, N_{1}, N_{2}, N_{m})$, where $U(S, N_{1}, N_{2}, N_{m})$ denotes the internal energy of the system and assume that the volumes are constant and the system is isolated. We denote by $\mu ^{k}= \frac{\partial U}{\partial N_k}$ the chemical potential of the chemical species in the reservoirs $(k=1,2)$ and in the membrane $(k=m)$. We denote by $\mathcal{J}^{1\rightarrow m}$ the flux from the reservoir $1$ into the membrane and $\mathcal{J}^{m\rightarrow 2}$ the flux from the membrane into the reservoir $2$. 

The variational condition for the diffusion process is provided by
\begin{equation}\label{var_cond_matter} 
\delta \int_{ t _1 }^{ t _2 }\left[ L(S, N_{1}, N_{2}, N_{m})+\dot{W}^{1} N_{1}+\dot{W}^{2} N_{2}+\dot{W}^{m} N_{m} \right]{\rm d}t =0,
\end{equation}
subject to the phenomenological constraint
\begin{equation}\label{phenom_constraint_matter} 
\frac{\partial L}{\partial S}\dot S   = \mathcal{J}^{m\rightarrow 1}(\dot W^{1} - \dot W^{m})  +\mathcal{J}^{m\rightarrow 2}(\dot W^{2} - \dot W^{m})
\end{equation} 
and for variations subject to the variational constraint
\begin{equation}\label{var_constraint_matter} 
\frac{\partial L}{\partial S}\delta  S   = \mathcal{J}^{m \rightarrow 1}( \delta  W^{1} - \delta  W^{m)}  +\mathcal{J}^{m\rightarrow 2}(\delta  W^{2} - \delta  W^{m}),
\end{equation}
with $\delta W^{k}( t _i )=0$, for $k=1,2,m$ and $i=1,2$.
\medskip

Thus, it follows
\begin{equation}\label{N_dot_equ_diff}
\dot N_{1} = \mathcal{J}^{m\rightarrow 1},\quad \dot N_{m} = \mathcal{J}^{1\rightarrow m}+\mathcal{J}^{2\rightarrow m}, \quad \dot N_{2} =\mathcal{J}^{m\rightarrow 2}
\end{equation} 
and $\dot W^{1} =\mu^{1},\;\dot W^{2} = \mu^{2},\; \dot W^{m}= \mu^{m}$. The constraint in \eqref{phenom_constraint_matter} becomes
\begin{equation}\label{S_dot_equ_diff}
-T\dot S= \mathcal{J}^{m\rightarrow 1}(\mu^{1}- \mu^{m})+ \mathcal{J}^{m\rightarrow 2}( \mu^{2}- \mu^{m}),
\end{equation}
where $T=-\frac{\partial L}{\partial S}$.
Equations \eqref{N_dot_equ_diff} and \eqref{S_dot_equ_diff} are equivalent with those derived in \cite[\S2.2]{OsPeKa1973}. From the equations \eqref{N_dot_equ_diff} and \eqref{S_dot_equ_diff}, we have the energy conservation $\frac{d}{dt}U= 0$, which is consistent with the fact that the system is isolated.

\medskip

\subsection{Adiabatically closed non-simple thermodynamic systems}\label{subsec_nonsimple}

\begin{figure}[h]
\begin{center}
\includegraphics[scale=0.55]{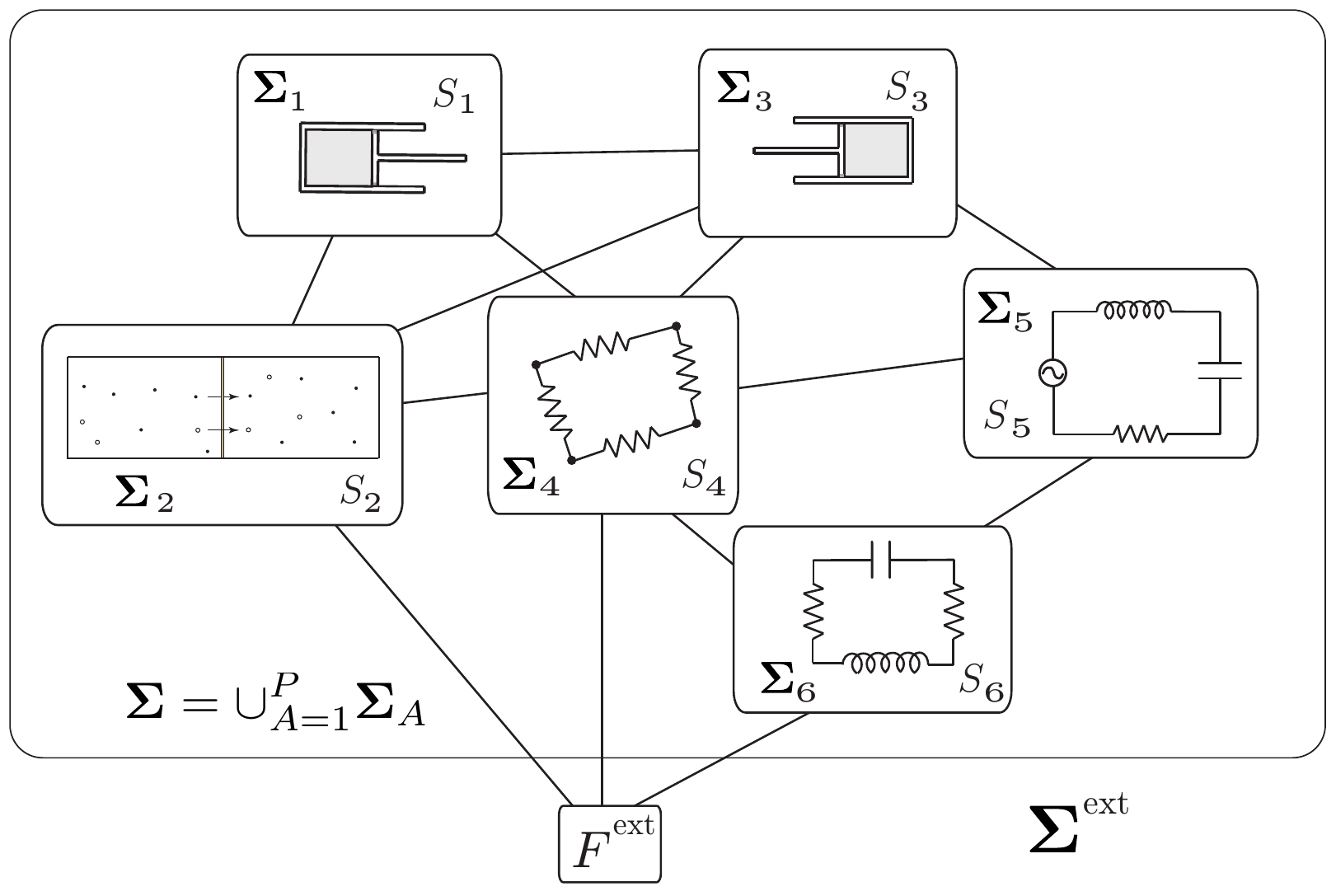}
\caption{Non-simple interconnected system.}
\label{Non-simple interconnected system}
\end{center}
\end{figure}

We now consider a general finite dimensional system $\boldsymbol{\Sigma}=\cup_{A=1}^P\boldsymbol{\Sigma}_A$, composed of interconnected simple thermodynamic systems $\boldsymbol{\Sigma}_A$, as illustrated in Fig.\,\ref{Non-simple interconnected system}. This class of non-simple interconnected systems extends the class of {\it interconnected mechanical systems} (see, \cite{JaYo2014}) to include the irreversible processes. In addition to the irreversible processes of friction and mass transfer described earlier, these systems can also involve the process of heat conduction.

\medskip

The main difference with the previous cases is the occurrence of several entropy variables, namely, each subsystem $\boldsymbol{\Sigma}_A$ has an entropy denoted $S_A,\,A=1,...,P$. Besides the variables $S_A$, each subsystem $\boldsymbol{\Sigma}_A$ may also be described by mechanical variables $q^A\in Q_A$ and number of moles $(N_{A,1},...,N_{A,K_{A}}) \in \mathbb{R}^{K_A}$, where $Q_A$ are the configuration manifolds for the mechanical variables associated to $\boldsymbol{\Sigma}_A$ and where $K_A$ are the numbers of compartments in the simple systems $\boldsymbol{\Sigma}_A$. For simplicity, we shall assume that independent mechanical coordinates $q\in Q$ have been chosen to represent the mechanical configuration of the interconnected system $\boldsymbol{\Sigma}$.
The state variables needed to describe this system are
\begin{equation}\label{list_variables}
(q,v)\in TQ,\qquad  S_A, \; A=1,...,P,\qquad N_{A,k}, \; k=1,...,K_A,\; A=1,...,P.
\end{equation}

We shall present the variational formulation for these systems in two steps, exactly as in \S\ref{subsec_simple} by first considering the case without any transfer of mass. 

\paragraph{(A) Variational formulation for systems with friction and heat conduction.} Beside the entropies $S_A$, $A=1,...,P$, these systems only involve mechanical variables. The Lagrangian of the system is thus a function
\[
L: TQ\ \times \mathbb{R}^P  \rightarrow \mathbb{R} , \quad \left(q, v,S_1,...,S_P\right) \mapsto L\left(q, v,S_1,...,S_P\right).
\]
We denote by $F^{{\rm ext}\rightarrow A}:T^*Q \times\mathbb{R}^P \rightarrow T^*Q$ the external force acting on subsystem $\boldsymbol{\Sigma}_A$. Consistently with the fact that the mechanical variables $q=(q^{1},...,q^{n})$ describe the configuration of the entire interconnected system $\boldsymbol{\Sigma}$, only the total exterior force $F^{\rm ext}=\sum_{A=1}^P F^{{\rm ext}\rightarrow A}$ appears explicitly in the variational condition \eqref{VCond_conduction}. We denote by $F^{{\rm fr} (A)}:T^*Q \times\mathbb{R}^P \rightarrow T^*Q$ the friction forces experienced by subsystem $\boldsymbol{\Sigma}_A$. This friction force is at the origin of an entropy production for subsystem $\boldsymbol{\Sigma}_A$ and appears explicitly in the phenomenological constraint \eqref{PC_conduction} and the variational constraint \eqref{VC_conduction} of the variational formulation. We also introduce the fluxes $J_{AB}$,  $A\neq B$ associated to the heat exchange between subsystems $\boldsymbol{\Sigma}_A$ and $\boldsymbol{\Sigma}_B$ and such that $J_{AB}=J_{BA}$. The relation between the fluxes $J_{AB}$ and the heat power exchange $P^{A\rightarrow B}_H$ is given later. For the construction of variational structures, it is convenient to define the flux $J_{AB}$ for $A=B$ as
\[
J_{AA}:=- \sum_{B\neq A}J_{AB},
\]
so that we have 
\begin{equation}\label{propertyJ}
\sum_{A=1}^PJ_{AB}=0,\quad\text{for all $B$}.
\end{equation}

\paragraph{Thermodynamic displacements associated to heat exchange.}  To incorporate heat exchange into our variational formulation, the new variables $\Gamma ^A$, $A=1,...,P$ are introduced. These are again examples of \textit{thermodynamic displacements} in the same way as we defined $W^k$ before. For the case of heat exchange, $\dot \Gamma^A$ corresponds to the temperature of the subsystem $\boldsymbol{\Sigma}_A$, where $\Gamma^A$ is identical to the \textit{thermal displacement} employed in \cite{GrNa1991}, which was originally introduced by \cite{He1884}. The introduction of $\Gamma^A$ is accompanied with the introduction of an entropy variable $\Sigma_A$ whose meaning will be clarified later.
\medskip

Now, the variational formulation for a system with friction and heat conduction is stated as follows:

\begin{framed}
\noindent Find the curves $q(t)$, $S_A(t)$, $\Gamma^A(t)$, $\Sigma_A(t)$ which are critical for the \textit{variational condition}
\begin{equation}\label{VCond_conduction}
\delta \int_{t _1 }^{ t _2} \!\Big[ L\left(q, \dot q, S_1,...,S_{K}\right)+  \dot{\Gamma }^A( S_A- \Sigma  _A)\Big] {\rm d}t + \int_{t_1}^{t_2}  \left<F^{\rm ext }, \delta q\right> \,{\rm d}t =0,
\end{equation}
subject to the \textit{phenomenological constraint}
\begin{equation}\label{PC_conduction}
\frac{\partial L}{\partial S_A}\dot \Sigma_A  =  \left<F^{{\rm fr}(A)}, \dot q \right>   + J_{AB}\dot\Gamma^B, \quad \text{for $A=1,...,P$},
\end{equation}
and for variations subject to the \textit{variational constraint}
\begin{equation}\label{VC_conduction}
\frac{\partial L}{\partial S_A}\delta \Sigma_A  = \left< F^{{\rm fr}(A)}, \delta q \right>   +  J_{AB}\delta\Gamma^B, \quad \text{for $A=1,...,P$},
\end{equation}
with $\delta q(t_1)=\delta q(t_2)=0$ and $ \delta \Gamma^A(t_1)=\delta \Gamma^A(t_2)=0$, $A=1,...,P$.
\end{framed}

\medskip

Taking variations of the integral in \eqref{VCond_conduction}, integrating by parts, and using $ \delta q(t_1)=\delta (t_2)=0$ and $ \delta \Gamma_A(t_1)=\delta \Gamma_A(t_2)=0$, it follows
\[
\int_{t_1}^{t_2}\left[\left( \frac{\partial L}{\partial q^i}-\frac{d}{dt} \frac{\partial L}{\partial \dot q^{i}}+F^{\rm ext }_i \right) \delta q^{ i}+ \frac{\partial L}{\partial S_A}\delta S_A - (\dot S_A- \dot \Sigma  _A)\delta\Gamma ^A + \dot \Gamma ^A( \delta S_A- \delta \Sigma  _A)
\right] {\rm d}t=0.
\]
Then, using the variational constraint \eqref{VC_conduction}, we get the following conditions:
\begin{align*} 
\delta q^{i}:&\quad\frac{\partial L}{\partial q^{i}}-\frac{d}{dt} \frac{\partial L}{\partial \dot q^{i}} -\sum_{A=1}^P \frac{\dot \Gamma^A}{ \frac{\partial L}{\partial S_A} }F^{{\rm fr}(A)}_{i} + F^{\rm ext }_{i}=0,\quad i=1,...,n,\\
\delta S_A:&\quad  \frac{\partial L}{\partial S_A}+ \dot\Gamma ^A=0, \quad A=1,...,P,\\
\displaystyle\delta \Gamma ^A: &\quad- \dot S_A + \dot \Sigma  _A -\sum_{B=1}^P \frac{\dot \Gamma ^A}{ \frac{\partial L}{\partial S_A} }J_{BA}=0,\quad A=1,...,P.
\end{align*}
The second equation yields
\begin{equation}\label{Gamma_T}
\dot\Gamma ^A= -\frac{\partial L}{\partial S_A}=:T^A,
\end{equation}
where $T^A$ is the temperature of the subsystem $\boldsymbol{\Sigma}_A$. This implies that $\Gamma_A$ is a thermal displacement. 
Because of \eqref{propertyJ}, the last equation yields $\dot S_A = \dot \Sigma  _A$. Hence, using \eqref{PC_conduction}, we get the following system of evolution equations for the curves $q(t)$ and $S_A(t)$:
\begin{equation}\label{systems_conduction} 
\left\{
\begin{array}{l}
\displaystyle\vspace{0.2cm}\frac{d}{dt}\frac{\partial L}{\partial \dot q}- \frac{\partial L}{\partial q}= \sum_{A=1}^P F^{{\rm fr}(A)} +F^{\rm ext} ,\\
\displaystyle\frac{\partial L}{\partial S_A}\dot S_A=  \left<F^{{\rm fr}(A)},\dot q \right> - \sum_{B=1}^P J_{AB}\left(\frac{\partial L}{\partial S_B} - \frac{\partial L}{\partial S_A}\right),\quad A=1,...,P .
\end{array} \right.
\end{equation} 

As before, we have $\frac{d}{dt}E= \left<F^{\rm ext}, \dot q \right>= P_W^{\rm ext}$, where the total energy $E$ is defined in the same way as before. Since the system is {\it non-simple}, it is instructive to analyze the energy behavior of each subsystem. This can be done if the Lagrangian is given by the sum of the Lagrangians of the subsystems, i.e.,
\[
L(q, v, S_1,...,S_P)= \sum_{A=1}^P L_A(q,v, S_A).
\]
The mechanical equation for $\boldsymbol{\Sigma}_A$ is given as
\[
\frac{d}{dt}\frac{\partial L_A}{\partial \dot q}- \frac{\partial L_A}{\partial q}=  F^{{\rm fr}(A)} +F^{{\rm ext}\rightarrow A}+\sum_{B=1}^PF^{B \rightarrow A},
\]
where $F^{B\rightarrow A}=-F^{A\rightarrow B}$ is the internal force exerted by $\boldsymbol{\Sigma}_B$ on $\boldsymbol{\Sigma}_A$. Denoting $E_A$ the total energy of $\boldsymbol{\Sigma}_A$, we have
\begin{equation}\label{detailed_energy_bal}\begin{aligned}
\frac{d}{dt}E_A &= \left<F^{{\rm ext}\rightarrow A}, \dot q \right> + \sum_{B=1}^P \left<F^{B \rightarrow A}, \dot q \right> + \sum_{B=1}^P J_{AB}\left(\frac{\partial L}{\partial S_B} - \frac{\partial L}{\partial S_A}\right) \\
&= P_W^{{\rm ext}\rightarrow A} +\sum_{B=1}^P P^{B \rightarrow A}_W + \sum_{B=1}^P P_H^{B \rightarrow A},
\end{aligned}
\end{equation}
where $P_W^{{\rm ext}\rightarrow A}$ and $P^{B \rightarrow A}_W$ denote the power associated to the  work done on $\boldsymbol{\Sigma}_A$ by the exterior and that by the subsystem $\boldsymbol{\Sigma}_B$ respectively, and where $P_H^{B \rightarrow A}$ is the power associated to the heat transfer from $\boldsymbol{\Sigma}_B$ to $\boldsymbol{\Sigma}_A$. The link between the flux $J_{AB}$ and the power exchange is thus
\[
P_H^{B \rightarrow A}= J_{AB}(T^A-T^B).
\]

Since entropy is an extensive variable, the total entropy of the system is $S=\sum_{A=1}^PS_A$. From \eqref{systems_conduction}, it follows that the rate of total entropy production $\dot{S}=\sum_{A=1}^{P}\dot S_A$ of the system is given by
\begin{equation}\label{internal_entropy_production}
\dot{S}=-\sum_{A=1}^P\frac{1}{T^A} \left<F^{{\rm fr}(A)}, \dot q \right> +
\sum_{A<B}^{K}J_{AB}\left(\frac{1}{T^B}-\frac{1}{T^A} \right) (T^B-T^A).
\end{equation}
The second law suggests the phenomenological relations
\begin{equation}\label{phen_rela}
F^{{\rm fr}(A)}_{i}=-\lambda^A_{ij} \dot{q}^{j}\qquad\text{and}\qquad J_{AB}\frac{T^A-T^B}{T^AT^B}=\mathcal{L}_{AB}(T^B-T^A),
\end{equation}
where $\lambda^A_{ij}$ and $\mathcal{L}_{AB}$ are functions of the state variables, with the symmetric part of the matrices $\lambda^A_{ij}$ positive semi-definite and with $\mathcal{L}_{AB}\geq 0$, for all $A,B$. From the second relation, we deduce $J_{AB}= -\mathcal{L}_{AB}T^AT^B=-\kappa_{AB}$, with $\kappa_{AB}=\kappa_{AB}(q, S_A, S_B)$ the heat conduction coefficients between subsystem $\boldsymbol{\Sigma}_A$ and subsystem $\boldsymbol{\Sigma}_B$.

\begin{figure}[h]
\begin{center}
\includegraphics[scale=0.6]{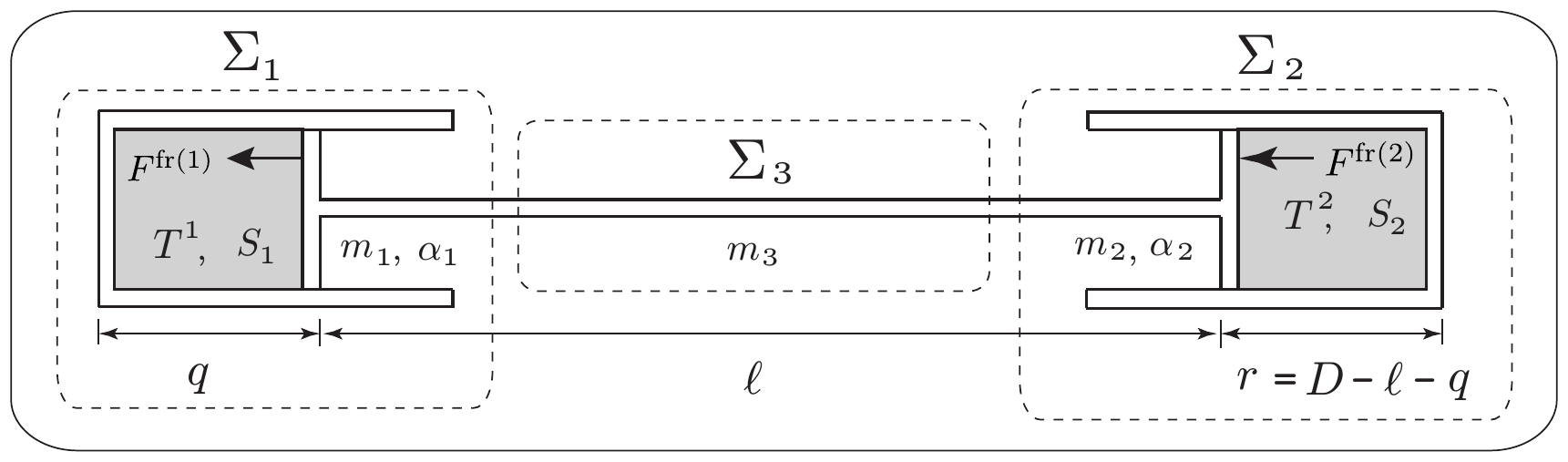}
\caption{The two-cylinder problem.}
\label{two_pistons}
\end{center}
\end{figure}

\paragraph{Example: the adiabatic piston.}\label{section-two-piston}
We consider a piston-cylinder system composed of two cylinders connected by a rod, each of which contains a fluid (or an ideal gas) and is separated by a movable piston, as shown in Fig.\,\ref{two_pistons}. We assume that the system is isolated. Despite its apparent simplicity, this system has attracted a lot of attention in the literature because there has been some controversy about the final equilibrium state of this system when the piston is adiabatic. We refer to \cite{Gr1999} for a review of this challenging problem and for the derivation of the time evolution of this system, based on the approach of \cite{StSc1974}. 
The system $ \boldsymbol{\Sigma} $ may be regarded as an interconnected system consisting of three simple systems; namely, the two pistons $ \boldsymbol{\Sigma} _1 , \boldsymbol{\Sigma} _2 $ of mass $ m_1 , m _2 $ and the connecting rod $ \boldsymbol{\Sigma} _3 $ of mass $ m _3 $. As illustrated in Fig.\,\ref{two_pistons}, $q$ and $r=D-\ell-q$ denote respectively the distance between the bottom of each piston to the top, where $D$ is a constant.
In this setting, we choose the variables $(q, v , S_1, S_2)$ (the entropy associated to $ \boldsymbol{\Sigma} _3 $ is constant) to describe the dynamics of the interconnected system and the Lagrangian is given by
\begin{equation}\label{Lagrangian_2_pistons}
L(q, v, S _1, S _2  )= \frac{1}{2} M v ^2 -U_1 (q, S _1 )-U_2(q, S _2 ),
\end{equation} 
where $M:= m_1  + m _2 + m _3$, and 
\[
U_1(q,S_1):=\mathsf{U}_1(S_1, V_1= \alpha_1 q, N_1),\quad U_2(q, S_2):= \mathsf{U}_2(S_2, V_2= \alpha_2 r, N_2),
\]
with $\mathsf{U}_i(S_i, V_i, N_i)$, the internal energies of the fluids, $N_i$ the constant number of moles, and $\alpha_i$ are the constant areas of the cylinders, $i=1,2$.

As in \eqref{phen_rela} we have $F^{{\rm fr} (A)}(q, \dot q,  S_A )= - \lambda ^A \dot q$, with $\lambda^A=\lambda^A(q,S^A)\geq 0$, $A=1,2$ and $J_{AB}=-\kappa_{AB}=:-\kappa$, where $\kappa=\kappa(S_1,S_2,q)\geq 0$ is the heat conductivity of the connecting rod.

From the variational formulation \eqref{VCond_conduction}--\eqref{VC_conduction},  we get the following system for $q(t)$, $S_1(t)$, $S_2(t)$, in view of \eqref{systems_conduction}, as
\[
\left\{ 
\begin{array}{l}
\displaystyle\vspace{0.2cm}M\ddot q= p _1 (q,S_1)\alpha_1 - p _2(q,S_2) \alpha_2 - (\lambda ^1 + \lambda ^2 )\dot q,\\
\displaystyle\vspace{0.2cm} T^1(q,S_1)\dot S_1= \lambda ^1 \dot q^2  +\kappa \left( T ^2(q,S_2)-T^1(q,S_1)\right),\\
\displaystyle T^2(q,S_2)\dot S_2= \lambda ^2\dot q^2  +\kappa \left(T^1(q,S_1)-T^2(q,S_2)\right),
\end{array} 
\right.
\]
where we used $\frac{\partial U_i}{\partial S_i}(q,S_i)=T^i(q,S_i)$, $\frac{\partial U_1 }{\partial q}=-p _1 (q, S_1)\alpha_1 $, and $ \frac{\partial U_2}{\partial q}=p _2(q, S_2) \alpha_2 $.

These equations recover those derived in \cite{Gr1999}, (51)--(53). We have $ \frac{d}{dt} E=0$, where $E= \frac{1}{2} M\dot q^2 +U_1(q,S_1)+U(q, S_2 )$, consistently with the fact that the system is isolated. The rate of total entropy production is
\[
\frac{d}{dt} S= \left( \frac{\lambda ^1}{T ^1} + \frac{\lambda ^2}{T ^2}  \right) \dot q^2 + \kappa\frac{(T ^2 - T ^1 ) ^2 }{T ^1 T ^2 } \geq 0.
\]
The equations of motion for the adiabatic piston are obtained by setting $ \kappa =0$.

\paragraph{(B) Variational formulation for systems with friction, heat conduction, and internal mass transfer.} We extend the previous case to the case in which the subsystems $\boldsymbol{\Sigma}_A$ not only exchange work and heat, but also matter. In general, each subsystem may itself have several compartments, in which case the variables are those listed in \eqref{list_variables}. For simplicity, we shall assume that each subsystem has only one compartment. The reader can easily extend this approach to the general case. The Lagrangian is thus a function
\[
L: TQ\times\mathbb{R}^P \times \mathbb{R}^P  \rightarrow \mathbb{R} , \quad \left(q, v,S_1,...,S_P, N_1,...,N_P\right) \mapsto L\left(q, v,S_1,...,S_P, N_1,...,N_P\right),
\]
where $S_A$ and $N_A$ are the entropy and number of moles of subsystem $\boldsymbol{\Sigma}_A$, $A=1,...,P$.
Since the previous cases have been already presented in details earlier, we shall here just present the variational formulation and the resulting equations of motion.

\begin{framed}
\noindent Find the curves $q(t)$, $S_A(t)$, $\Gamma^A(t)$, $\Sigma_A(t)$, $W^A(t)$, $N_A(t)$ which are critical for the \textit{variational condition}
\begin{equation}\label{VCond_conduction_matter}
\begin{aligned}
&\delta \int_{t _1 }^{ t _2} \!\Big[ L\left(q, \dot q, S_1,...,S_P, N_1,...,N_P\right) + \dot W^A N_A+  \dot{\Gamma }^A( S_A- \Sigma  _A)\Big] {\rm d}t+ \int_{t_1}^{t_2}  \left<F^{\rm ext }, \delta q \right>{\rm d}t =0,
\end{aligned}
\end{equation}
subject to the \textit{phenomenological constraint}
\begin{equation}\label{PC_conduction_matter}
\frac{\partial L}{\partial S_A}\dot \Sigma_A  =  \left<F^{{\rm fr}(A)}, \dot q\right>   + J_{AB}\dot\Gamma^B + \mathcal{J}^{B \rightarrow A}\dot W^A, \quad \text{for $A=1,...,P$},
\end{equation}
and for variations subject to the \textit{variational constraint}
\begin{equation}\label{VC_conduction_matter}
\frac{\partial L}{\partial S_A}\delta \Sigma_A  =  \left<F^{{\rm fr}(A)},\,  \delta q\right>   +  J_{AB}\delta\Gamma^B+ \mathcal{J}^{B \rightarrow A}\delta W^A, \quad \text{for $A=1,...,P$},
\end{equation}
with $\delta q(t_i)=\delta W^A(t_i)=\delta \Gamma^A(t_i)=0$, $i=1,2$, $A=1,...,P$.
\end{framed}

\medskip

From \eqref{VCond_conduction_matter}--\eqref{VC_conduction_matter}, we obtain the following system of evolution equations for the curves $q(t)$, $S_A(t)$, and $N_A(t)$:
\begin{equation}\label{systems_conduction_diffusion} 
\left\{
\begin{array}{l}
\displaystyle\vspace{0.2cm}\frac{d}{dt}\frac{\partial L}{\partial \dot q}- \frac{\partial L}{\partial q}= \sum_{A=1}^P F^{{\rm fr}(A)} +F^{\rm ext},\\
\displaystyle\vspace{0.2cm}\frac{d}{dt} N_A= \sum_{B=1}^{P} \mathcal{J}^{B \rightarrow A},\quad A=1,...,P,\\
\displaystyle\vspace{0.2cm}\frac{\partial L}{\partial S_A}\dot S_A=  \left<F^{{\rm fr}(A)}, \dot q\right> - \sum_{B=1}^P J_{AB}\left(\frac{\partial L}{\partial S_B} - \frac{\partial L}{\partial S_A}\right)-\sum_{B=1}^P \mathcal{J} ^{B \rightarrow A} \frac{\partial L}{\partial N_A}, \quad A=1,...,P.
\end{array} \right.
\end{equation} 
We also obtain the conditions
\[
\dot\Gamma ^A=-\frac{\partial L}{\partial S_A}=:T^A, \qquad \dot W^A=-\frac{\partial L}{\partial N_A}=: \mu ^A,\qquad \dot\Sigma _A= \dot S_A , \quad A=1,...,P,
\]
where we defined the temperature $T^A$ and the chemical potential $\mu^A$ of the subsystem $\boldsymbol{\Sigma}_A$. The variables $\Gamma^A$ and $W^A$ are again the thermodynamic displacements associated to the processes of heat and matter transfer.

\medskip

The total energy satisfies $\frac{d}{dt}E= P_W^{\rm ext}$ and the detailed energy balances can be carried out as in \eqref{detailed_energy_bal} and yields here
\[
P^{B\rightarrow A}_{H+M}= J_{AB}(T^A- T^B).
\]
The rate of total entropy production of the system is computed as
\[
\dot{S}=-\sum_{A=1}^{P}\frac{1}{T^A} \left<F^{{\rm fr}(A)}, \dot q \right> +
\sum_{A<B}J_{AB}\left(\frac{1}{T^B}-\frac{1}{T^A} \right)(T^B-T^A)
+\sum_{A<B}\mathcal{J}^{B \rightarrow A}\left(\frac{\mu^B}{T^B}-\frac{\mu^A}{T^A} \right).
\]
From the second law of thermodynamics, the total entropy production must be positive and hence suggests the phenomenological relations

\begin{equation}\label{pheno_open_system}
F^{{\rm fr}(A)}_{i}=-\lambda^A_{ij} \dot{q}^{j},\qquad\qquad
\begin{bmatrix}
\vspace{0.2cm}\frac{T^A-T^B}{T^AT^B}J_{AB}\\
\mathcal{J}^{B \rightarrow A}
\end{bmatrix}
=
\mathcal{L}_{AB}
\begin{bmatrix}
\vspace{0.2cm} T^B - T^A \\
\frac{\mu^B}{T^B}-\frac{\mu^A}{T^A}
\end{bmatrix},
\end{equation}
where the symmetric part of the $n\times n$ matrices $\lambda^A$ and of the $2\times 2$ matrices $\mathcal{L}_{AB}$ are positive.
The entries of these matrices are phenomenological coefficients determined experimentally, which may in general depend on the state variables. From Onsager's reciprocal relations, the $2\times 2$ matrix
\[
\mathcal{L}_{AB}=\begin{bmatrix}
\vspace{0.3cm}\mathcal{L}_{AB}^{HH} & \mathcal{L}_{AB}^{HM}\\
\mathcal{L}_{AB}^{MH}& \mathcal{L}_{AB}^{MM}
\end{bmatrix}
\]
is symmetric, for all $A,B$. The matrix elements $\mathcal{L}_{AB}^{HH} $ and $\mathcal{L}_{AB}^{MM}$ are related to the processes of heat conduction and diffusion between $\boldsymbol{\Sigma}_A$ and $\boldsymbol{\Sigma}_B$. The coefficients $\mathcal{L}_{AB}^{MH}$ and $\mathcal{L}_{AB}^{HM}$ describe the cross-effects, and hence are associated to discrete versions of the process of thermal diffusion and the Dufour effect. Thermal diffusion is the process of matter diffusion due to the temperature difference between the compartments. The Dufour effect is the process of heat transfer due to difference of chemical potentials between the compartments.
\medskip

\begin{figure}[h]
\begin{center}
\includegraphics[scale=0.6]{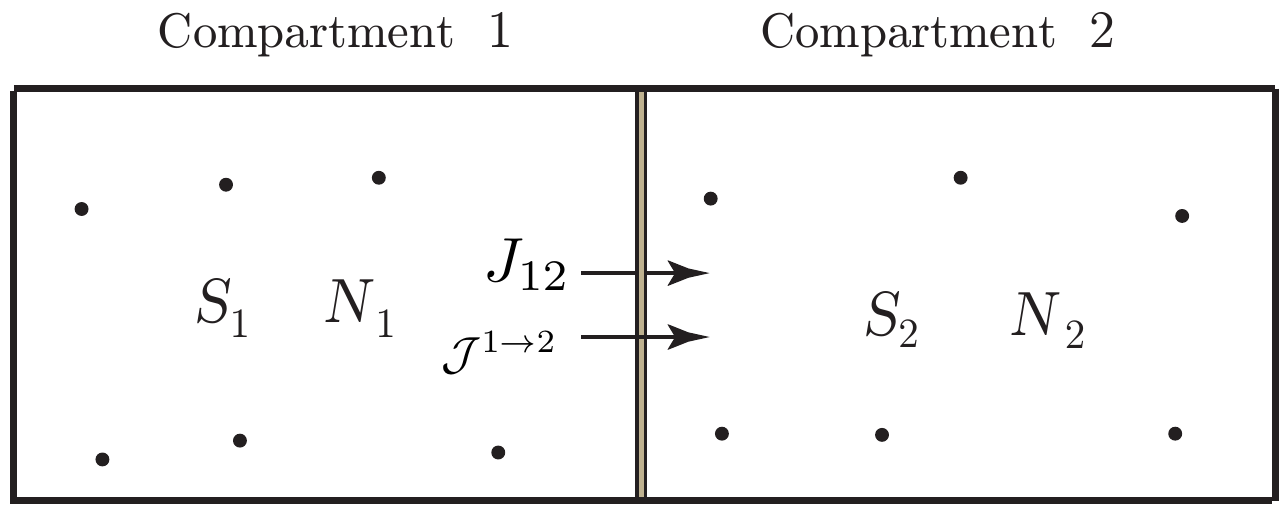}
\caption{Non-simple closed system with a single chemical species, experiencing diffusion and heat conduction between two compartments.}
\label{ClosedMassHeatTransSim}
\end{center}
\end{figure}

\noindent\textbf{Example: heat conduction and diffusion between two compartments.} We consider a closed system consisting of two compartments as illustrated in Fig.\,\ref{ClosedMassHeatTransSim}. The compartments are separated by a permeable wall through which heat conduction and diffusion is possible. The system is closed and therefore there is no matter transfer with exterior, while we have heat and mass transfer between the compartments. 

The Lagrangian of this system is
\[
L(S_1,S_2,N_1, N_2)=- U_{1}(S_1,N_1) - U_{2}(S_2,N_2) ,
\]
where $U_{i}(S_{i}, N_{i})$ is the internal energy of the $i$-th chemical species and the volume is assumed to be constant.
In this case, the system \eqref{systems_conduction_diffusion} specifies to
\begin{equation}\label{SimpleHeatMassTransferEqn}
\left\{ 
\begin{array}{l}
\displaystyle \dot N_1=  \mathcal{J} ^{ 2 \rightarrow 1},\qquad \dot N_2=  \mathcal{J} ^{ 1 \rightarrow 2},\\[3mm]
\displaystyle T^{1}\dot S_{1}=-J^{12}(T^{2}-T^{1})-\mathcal{J} ^{2 \rightarrow 1} \mu^{1} ,\\[3mm]
\displaystyle T^{2}\dot S_{2}=-J^{12}(T^{1}-T^{2})-\mathcal{J} ^{1 \rightarrow 2} \mu^{2},
\end{array}
\right.
\end{equation} 
where 
\[
T^{A}=\frac{\partial U}{\partial S_{A}},\qquad \mu^{A}=\frac{\partial U}{\partial N_{A}},\;\;A=1,2
\]
are the temperatures and chemical potentials of the $A$-th compartments.
From \eqref{SimpleHeatMassTransferEqn} it follows that the equation for the total entropy $S=S_1+S_2$ of the system is
\[
\dot S
=J^{12}(T^{1}-T^{2})\left(\frac{1}{T^{1}}-\frac{1}{T^{2}}\right)+\mathcal{J}^{1 \rightarrow 2} \left(\frac{\mu^{1}}{T^{1}}-\frac{\mu^{2}}{T^{2}}\right)\geq 0,
\]
from which the phenomenological relations are obtained as in \eqref{pheno_open_system}. The energy balance in each compartment is
\[
\frac{d}{dt}U_1=-J^{12}(T^2-T^1),\qquad \frac{d}{dt}U_2=-J^{12}(T^1-T^2),
\]
which shows the relation between the flux $J^{12}$ and the power $P^{1\rightarrow 2}=J^{12}(T^2-T^1)$ exchanged between the two compartments, due to heat conduction, diffusion, and their cross-effects.  The total energy $E=U_{1}+U_{2}$ is conserved.

\begin{remark}[General structure of the variational formulation for adiabatically closed systems]\label{PC_VS_VC_general}{\rm In each of the situation considered, the variational constraint can be systematically obtained from the phenomenological constraint by replacing the time derivative by the delta variation, for each process. For the most general case treated above, we have the following correspondence:
\begin{align*}
&\frac{\partial L}{\partial S_A}\dot \Sigma_A  =  \left<F^{{\rm fr}(A)},\dot q \right>   + J_{AB}\dot\Gamma^B + \mathcal{J}^{B \rightarrow A}\dot W^A\quad\leadsto \quad \frac{\partial L}{\partial S_A}\delta \Sigma_A  =  \left<F^{{\rm fr}(A)}, \delta q  \right> +  J_{AB}\delta\Gamma^B+ \mathcal{J}^{B \rightarrow A}\delta W^A.
\end{align*}
In the above, the quantities to be determined from the state variables by phenomenological laws are $F^{{\rm fr}(A)}$, $J_{AB}$, and $\mathcal{J}^{B \rightarrow A}$.

\medskip

The structure of our variational formulation is better explained by adopting a general point of view. 
If we denote by $ \mathcal{Q}$ the thermodynamic configuration manifold and by $x\in \mathcal{Q}$ the collection of all the variables of the thermodynamic system, for instance $x=(q,S_A, N_A, W^A, \Gamma^A,\Sigma_A)$, $A=1,...,P$ in the preceding case,  then the variational formulation for adiabatically closed system falls into the following abstract setting.
Given a Lagrangian $\mathcal{L}:T\mathcal{Q}\rightarrow\mathbb{R}$, an external force $\mathcal{F}^{\rm ext}:T\mathcal{Q}\rightarrow T^*\mathcal{Q}$, and fiber preserving maps $A^\alpha:T\mathcal{Q}\rightarrow T^*Q$, $A^\alpha(x,v)\in T^*_x\mathcal{Q}$, $\alpha=1,...,k$, the variational formulation reads as follows:
\begin{equation}\label{VCond_general}
\delta  \int_{t_1 }^{t_2 }\mathcal{L}(x(t), \dot x(t)) {\rm d}t +   \int_{t_1 }^{t_2 } \left<\mathcal{F}^{\rm ext}(x(t), \dot x(t)),\delta x(t)\right>{\rm d}t=0, 
\end{equation}
where the curve $x(t)$ satisfies the \textit{phenomenological constraint}
\begin{equation}\label{PC_general}
A^\alpha(x,\dot x)\!\cdot\!\dot x=0,\;\;\text{for $\alpha=1,...,k$},
\end{equation}
and for variations $\delta x$ subject to the \textit{variational constraint}
\begin{equation}\label{VC_general}
A^\alpha(x,\dot x)\!\cdot\!\delta x =0 ,\;\;\text{for $\alpha=1,...,k$},
\end{equation}
with $\delta x(t_1)=\delta x(t_2)=0$.
\medskip

This yields the system of equations
\begin{equation}\label{general_system}
\left\{
\begin{array}{l}
\displaystyle\vspace{0.2cm}\frac{d}{dt} \frac{\partial \mathcal{L}}{\partial \dot x}- \frac{\partial \mathcal{L}}{\partial x} - \mathcal{F}^{\rm ext}=\lambda_\alpha A^\alpha(x, \dot x),\\
\displaystyle A^\alpha(x, \dot x)\!\cdot\!\dot x=0,\;\;\alpha=1,...,k.
\end{array}\right.
\end{equation}
It is clear that all the variational formulations for adiabatically closed system considered above fall into this category, by appropriate choice for $x$, $\mathcal{L}(x, \dot x)$, $\mathcal{F}^{\rm ext}(x,\dot x)$, and $A^\alpha(x, \dot x)$. The energy defined by $\mathcal{E}(x,v)=\left<\frac{\partial \mathcal{L}}{\partial v}, v \right>- \mathcal{L}(x,v)$ satisfies $
\frac{d}{dt}\mathcal{E}= \left<\mathcal{F}^{\rm ext},\dot x\right>$.

\medskip

The constraints involved in this variational formulation admit an intrinsic geometric description. The variational constraint \eqref{VC_general} defines the subset $C_V\subset T\mathcal{Q}\times_\mathcal{Q} T\mathcal{Q}$ given by
\[
C_V=\{(x, v, \delta x)\in T\mathcal{Q}\times_\mathcal{Q} T\mathcal{Q}\mid A^\alpha(x,v)\!\cdot\!\delta x =0 ,\;\;\text{for $\alpha=1,...,k$}\},
\]
so that $C_V(x,v):= C_V\cap \big(\{(x,v)\}\times T_x\mathcal{Q}\big)$
is a vector subspace of $T_x\mathcal{Q}$ for all $(x,v)\in T\mathcal{Q}$. 
The phenomenological constraint \eqref{PC_general} defines the subset $C_K\subset T\mathcal{Q}$ given by
\[
C_K=\{(x,v)\in T\mathcal{Q}\mid A^\alpha(x,v)\!\cdot\!v =0 ,\;\;\text{for $\alpha=1,...,k$}\}.
\]
Then, one notes that the constraint $C_K$ can be {\it intrinsically defined} from $C_V$ as
\[
C_K= \{(x,v)\in T\mathcal{Q}\mid (x,v)\in C_V(x,v)\}.
\]
Constraints $C_V$ and $C_K$ related in this way are called \textit{nonlinear nonholonomic constraints of thermodynamic type}, see \cite{GBYo2017a,GBYo2018c}.}
\end{remark}

\subsection{Open thermodynamic systems}

The thermodynamic systems that we considered so far are restricted to the adiabatically closed cases. For such systems, the interaction with the exterior is only through the exchange of mechanical work, and hence the first law for such systems reads
\[
\frac{d}{dt}E= \left<F^{\rm ext}, \dot q \right>=P^{\rm ext}_W.
\]

\medskip

We now consider the more general case of open systems exchanging work, heat, and matter with the exterior. In this case, the first law reads  
\[
\frac{d}{dt} E = P^{\rm ext}_W+P^{\rm ext}_H+P^{\rm ext}_M,
\]
where $P^{\rm ext}_H$ is the power associated to the transfer of heat into the system and $P^{\rm ext}_M$ is the power associated to the transfer of matter into the system. As we recall below, to the transfer of matter into or out of the system is associated a transfer of work and heat. By convention, $ P^{\rm ext}_W$ and $P^{\rm ext}_H$ denote uniquely the power associated to work and heat that is not associated to a transfer of matter. The power associated to a transfer of heat or work due to a transfer of matter is included in $P^{\rm ext}_M$.

\medskip

In order to get a concrete expression for $P^{\rm ext}_M$, let us consider an open system with several ports, $a=1,...,A$, through which matter can flow into or out of the system. We suppose, for simplicity, that the system involves only one chemical species and denote by $N$ the number of moles of this species. The mole balance equation is
\[
\frac{d}{dt}N=\sum_{a=1}^A \mathcal{J}^a,
\]
where $\mathcal{J}^a$ is the molar flow rate \textit{into} the system through the $a$-th port, so that $\mathcal{J}^a>0$ for flow into the system and $\mathcal{J}^a<0$ for flow out of the system.

As matter enters or leaves the system, it carries its internal, potential, and kinetic energy. This energy flow rate at the $a$-th port is the product $\mathsf{E}^a\mathcal{J}^a$ of the energy per mole (or molar energy) $\mathsf{E}^a$ and the molar flow rate $\mathcal{J}^a$ at the $a$-th port. In addition, as matter enters or leaves the system it also exerts work on the system that is associated with pushing the species into or out of the system. The associated energy flow rate is given at the $a$-th port by $\mathcal{J}^ap^a\mathsf{V}^a$, where $p^a$ and $\mathsf{V}^a$ are the pressure and the molar volume of the substance flowing through the $a$-th port. From this we get the expression
\begin{equation}\label{law1_explicit}
P^{\rm ext}_M= \sum_{a=1}^A\mathcal{J}^a(\mathsf{E}^a+ p^a\mathsf{V}^a).
\end{equation}
We refer, for instance, to \cite{Sa2006}, \cite{KlNe2011} for the detailed explanations of the first law for open systems.

\medskip

We present below an extension of the variational formulation to the case of open systems. In order to motivate the form of the constraints that we use, we first consider a particular case of simple open system, namely, the case of a system with a single chemical species $N$ in a single compartment with constant volume $V$ and without mechanical effects.
In this particular situation, the energy of the system is given by the internal energy written as $U=U(S,N)$, since $V=V_0$ is constant. The balance of mole and energy are respectively given by
\[
\frac{d}{dt}N=\sum_{a=1}^A \mathcal{J}^a, \qquad \frac{d}{dt} U  = \sum_{a=1}^A\mathcal{J}^a(\mathsf{U}^a+ p^a\mathsf{V}^a)=\sum_{a=1}^A\mathcal{J}^a\mathsf{H}^a,
\]
see \eqref{law1_explicit}, where $\mathsf{H}^a=\mathsf{U}^a+ p^a\mathsf{V}^a$ is the molar enthalpy at the $a$-th port and where $\mathsf{U}^a$, $ p^a$, and $\mathsf{V}^a$ are respectively the molar internal energy, the pressure and the molar volume at the $a$-th port. From these equations and the second law, one obtains the equations for the rate of change of the entropy of the system as
\begin{equation}\label{S_dot_simple}
\frac{d}{dt}S= I+\sum_{a=1}^A \mathcal{J}^a\mathsf{S}^a,
\end{equation}
where $\mathsf{S}^a$ is the molar entropy at the $a$-th port and $I$ is the rate of internal entropy production of the system given by
\begin{equation}\label{I_simple_example}
I= \frac{1}{T}\sum_{a=1}^A \mathcal{J}^a\left(\mathsf{H}^a-T\mathsf{S}^a- \mu \right),
\end{equation}
with $T= \frac{\partial U}{\partial S}$ the temperature and $\mu=\frac{\partial U}{\partial N}$ the chemical potential.
For our variational treatment, it is useful to rewrite the rate of internal entropy production as
\[
I= \frac{1}{T}\sum_{a=1}^A \Big[\mathcal{J}^a_S(T^a-T)+\mathcal{J}^a(\mu^a- \mu )\Big],
\]
where we defined the entropy flow rate $\mathcal{J}^a_S:=\mathcal{J}^a\mathsf{S}^a$ and also used the relation $\mathsf{H}^a=\mathsf{U}^a+ p^a\mathsf{V}^a= \mu^a +T^a\mathsf{S}^a$. The thermodynamic quantities known at the $a$-th port are usually the pressure $p^a$  and the temperature $T^a$, from which the other thermodynamic quantities, such as $\mu^a=\mu^a(p^a,T^a)$ or $\mathsf{S}^a=\mathsf{S}^a(p^a,T^a)$ are deduced in view of the state equations of the gas.

\medskip

Here, we only show the variational formulation for a simplified case of open systems, namely, an open system with only one entropy variable and one compartment with a single species. So, the open system is a simple system.
The reader is referred to \cite{GBYo2018a} for the more general cases of open systems, such as the  extensions of \eqref{VCond_simple_diffusion}--\eqref{VC_simple_diffusion} and \eqref{VCond_conduction_matter}--\eqref{VC_conduction_matter} to open systems, as well as for the case when the mechanical energy of the species is taken into account.

\medskip

The state variables needed to describe the system are $(q,v,S,N)\in TQ$ and the Lagrangian is a map
\[
L: TQ\times\mathbb{R}  \times \mathbb{R}   \rightarrow \mathbb{R} , \quad (q, v,S, N) \mapsto L(q, v,S, N),
\]
We assume that the system has $A$ ports, through which species can flow out or into the system and $B$ heat sources. As above, $\mu^a$ and $T^a$ denote the chemical potential and temperature at the $a$-th port and $T^b$ denotes the temperature of the $b$-th heat source.

\begin{framed}
\noindent Find the curves $q(t)$, $S(t)$, $\Gamma(t)$, $\Sigma(t)$, $W(t)$, $N(t)$ which are critical for the \textit{variational condition}
\begin{equation}\label{VCond_open}
\delta \int_{t _1 }^{ t _2} \!\Big[ L(q, \dot q, S,N) + \dot W N+  \dot{\Gamma }( S- \Sigma )\Big] {\rm d}t + \int_{t_1}^{t_2}  \left<F^{\rm ext }, \delta q\right> {\rm d}t =0,
\end{equation}
subject to the \textit{phenomenological constraint}
\begin{equation}\label{PC_open}
\frac{\partial L}{\partial S}\dot \Sigma  =  \left<F^{\rm fr}, \dot q \right>  +\sum_{a=1}^A \Big[\mathcal{J}^a(\dot W- \mu^a) +\mathcal{J}_S^a(\dot \Gamma- T^a)\Big] + \sum_{b=1}^B\mathcal{J}_S^b(\dot \Gamma- T^b),
\end{equation}
and for variations subject to the \textit{variational constraint}
\begin{equation}\label{VC_open}
\frac{\partial L}{\partial S}\delta \Sigma  = \left< F^{\rm fr}, \delta q \right>  + \sum_{a=1}^A \Big[\mathcal{J}^a \delta W +\mathcal{J}_S^a \delta \Gamma\Big] + \sum_{b=1}^B\mathcal{J}_S^b \delta \Gamma,
\end{equation}
with $\delta q(t_1)=\delta q(t_2)=0$, $ \delta W(t_1)=\delta W(t_2)=0$, and $ \delta \Gamma(t_1)=\delta \Gamma(t_2)=0$.
\end{framed}

We note that the variational constraint \eqref{VC_open} follows from the phenomenological constraint \eqref{PC_open} by formally replacing the time derivatives $\dot\Sigma$, $\dot q$, $\dot W$, $\dot \Gamma$ by the corresponding virtual displacements $\delta\Sigma$, $\delta q$, $\delta W$, $\delta \Gamma$, and by removing all the terms that depend uniquely on the exterior, i.e., the terms $\mathcal{J} ^a\mu ^a$, $\mathcal{J} ^a_S T ^a $, and $ \mathcal{J} ^b_ST ^b $. Such a systematic correspondence between the phenomenological and variational constraints, extends to open system the correspondence for adiabatically closed systems verified in  \eqref{CK_simple} $\leadsto$ \eqref{CV_simple}, \eqref{PC_simple_diffusion} $\leadsto$ \eqref{VC_simple_diffusion}, \eqref{PC_conduction} $\leadsto$ \eqref{VC_conduction}, \eqref{PC_conduction_matter} $\leadsto$ \eqref{VC_conduction_matter}, see also Remarks \ref{PC_VS_VC} and \ref{PC_VS_VC_general}. Note that the action functional in \eqref{VCond_open} has the same form as that in the case of adiabatically closed systems.

\medskip

Taking variations of the integral in \eqref{VCond_open}, integrating by parts, and using $ \delta q(t_1)=\delta (t_2)=0$, $\delta W(t_1)=\delta W(t_2)=0$, and $\delta \Gamma(t_1)=\delta \Gamma(t_2)=0$ and using the variational constraint \eqref{VC_open}, we obtain the following conditions:
\begin{equation}\label{conditions_open}
\begin{aligned}
\delta q:&\quad\frac{d}{dt}\frac{\partial L}{\partial \dot q^{i}}- \frac{\partial L}{\partial  q^{i}} = F^{\rm fr}_i+F^{\rm ext}_i,\quad i=1,...,n, \\
\delta S:&\quad\dot \Gamma = -\frac{\partial L}{\partial S}, \\
\delta W:&\quad\dot N= \sum_{a=1}^A\mathcal{J} ^a, \\
\delta N:&\quad\dot W=  -\frac{\partial L}{\partial N}, \\
\delta \Gamma :&\quad\dot S=\dot \Sigma +\sum_{a=1}^A\mathcal{J}^a_S+\sum_{b=1}^B\mathcal{J}^b_S.
\end{aligned}
\end{equation}
By the second and fourth equations the variables $\Gamma$ and $W$ are thermodynamic displacements as before. The main difference with the earlier cases is that now $\dot S$ and $\dot\Sigma$ are no more equal. The physical interpretation of $\do\Sigma$ is given below. From \eqref{PC_open}, it follows the system of evolution equations for the curves $q(t)$, $S(t)$, $N(t)$:
\begin{equation}\label{open_system} 
\left\{
\begin{array}{l}
\displaystyle\vspace{0.2cm}\frac{d}{dt}\frac{\partial L}{\partial \dot q}- \frac{\partial L}{\partial q}= F^{\rm fr} +F^{\rm ext},\quad\quad\quad  \frac{d}{dt} N= \sum_{a=1}^A \mathcal{J}^a,\\
\displaystyle\vspace{0.2cm}\frac{\partial L}{\partial S}\Big(\dot S -\sum_{a=1}^A \mathcal{J}_S^a - \sum_{b=1}^B \mathcal{J}_S^b\Big)\\
\displaystyle\quad=  \left<F^{\rm fr}, \dot q \right> -\sum_{a=1}^A\left[\mathcal{J}^a\Big(\frac{\partial L}{\partial N}+\mu^a\Big)+\mathcal{J}^a_S\Big(\frac{\partial L}{\partial S}+ T^a\Big)\right]-\sum_{b=1}^B \mathcal{J}^b_S\Big(\frac{\partial L}{\partial S}+ T^b\Big).
\end{array} \right.
\end{equation} 
The energy balance for this system is computed as
\[
\frac{d}{dt}E=\underbrace{\left<F^{\rm ext},\dot q\right>}_{=P^{\rm ext}_W} + \underbrace{\sum_{b=1}^B \mathcal{J}^b_S T^b}_{=P^{\rm ext}_H}+ \underbrace{\sum_{a=1}^A (\mathcal{J}^a\mu^a + \mathcal{J}_S^a T^a)}_{=P^{\rm ext}_M} .
\]
From the last equation in \eqref{open_system}, the rate of entropy equation of the system is found as
\begin{equation}\label{entropy_open}
\dot S=I + \sum_{a=1}^A \mathcal{J}^a_S + \sum_{b=1}^B \mathcal{J}^b_S,
\end{equation}
where $I$ is the rate of internal entropy production given by
\[
I= \underbrace{-\frac{1}{T}\left<F^{\rm fr} , \dot q \right>}_{\text{mechanical friction}} +\underbrace{\frac{1}{T} \sum_{a=1}^A\left[\mathcal{J}^a\Big(\mu ^ a- \mu\Big)+\mathcal{J}^a_S\Big( T^a - T\Big)\right]}_{\text{mixing of matter flowing into the system}} + \underbrace{\frac{1}{T}\sum_{b=1}^B \mathcal{J}^b_S\Big(T^b- T\Big)}_{\text{heating}}.
\]
From the last equation \eqref{conditions_open} and \eqref{entropy_open} {\it we notice that $\dot\Sigma =  I$ is the rate of internal entropy production}.
The second and third terms in \eqref{entropy_open} represent the entropy flow rate into the system associated to the ports and the heat sources. The second law requires $I\geq 0$, whereas the sign of the rate of entropy flow into the system is arbitrary.

\paragraph{Example: A piston device with ports and heat sources.} 
We consider a piston with mass $m$ moving in a cylinder containing a species with internal energy $U(S,V,N)$. We assume that the cylinder has two external heat sources with entropy flow rates $\mathcal{J}^{b_i}$, $i=1,2$, and two ports through which the species is injected into or flows out of the cylinder with molar flow rates $\mathcal{J}^{a_i}$, $i=1,2$. The entropy flow rates at the ports are given by $\mathcal{J}_S^{a_i}=\mathcal{J}^{a_i}\mathsf{S}^{a_i}$.

\begin{figure}[h]
\begin{center}
\includegraphics[scale=0.95]{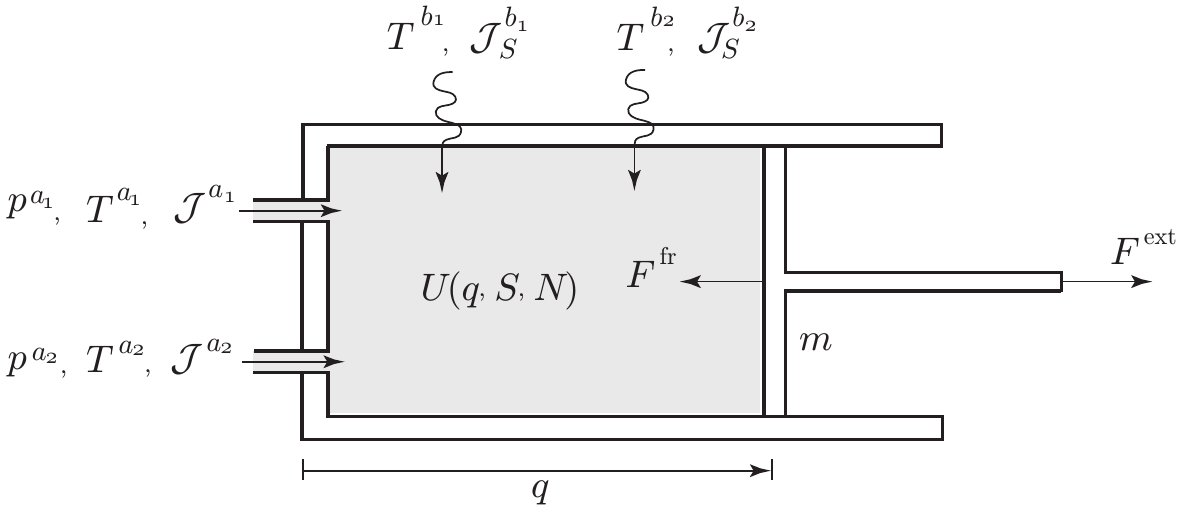}
\caption{A piston device with ports and heat sources.}
\label{flowing_piston}
\end{center}
\end{figure}

The variable $q$ characterizes the one-dimensional motion of the piston, such that the volume occupied by the species is $V=\alpha q$, with $\alpha$ the sectional area of the cylinder. The Lagrangian of the system is
\[
L(q, \dot q, S, N)= \frac{1}{2}m\dot q ^2 -U(S, Aq, N).
\]
The variational formulation \eqref{VCond_open}--\eqref{VC_open} yields the evolution equations for $q(t)$, $S(t)$, $N(t)$
\begin{equation*}\label{equation_motion_2}
m\ddot q =p(q,S,N)\alpha + F^{\rm fr}+F^{\rm ext}, \qquad   \dot N=\sum_{a=1}^A\mathcal{J}^a,\qquad \dot S=I+\sum_{i=1}^2\mathcal{J}_{S}^{a_{i}}+ \sum_{j=1}^2\mathcal{J}_{S}^{b_{j}},
\end{equation*}
where $p(q,S,N)=-\frac{\partial U}{\partial V}$ is the pressure and $I=\dot\Sigma$ is the internal entropy production given by
\[
I= - \frac{1}{T}F^{\rm fr}\dot q +\frac{1}{T}\sum_{i=1}^2 \Big[ (\mu^{a_{i}}-\mu)+\mathsf{S}^{a_{i}}(T^{a_{i}}-T)\Big]\mathcal{J}^{a_{i}}+ \frac{1}{T}\sum_{j=1}^2\mathcal{J}_S^{b_{j}}(T^{b_{j}}-T).
\]
The first term represents the entropy production associated to the friction experiencing by the moving piston, the second term is the entropy production associated to the mixing of gas flowing into the cylinder at the two ports $a_1$, $a_2$, and the third term denotes the entropy production due to the external heating.   
The second law requires that each of these terms is positive. The energy balance holds as
\[
\frac{d}{dt}E=\underbrace{F^{\rm ext}\dot q}_{=P^{\rm ext}_W} + \underbrace{\sum_{j=1}^2 \mathcal{J}^{b_{j}}_S T^{b_{j}}}_{=P^{\rm ext}_H}+ \underbrace{\sum_{i=1}^2 (\mathcal{J}^{a_{i}}\mu^{a_{i}} + \mathcal{J}_S^{a_{i}} T^{a_{i}})}_{=P^{\rm ext}_M}.
\]

\begin{remark}[Inclusion of chemical reactions]{\rm The variational formulations presented so far, can be extended to include several chemical species, undergoing chemical reactions.
Let us denote by $I=1,...,R$ the chemical species and by $a=1,...,r$ the chemical reactions. Chemical reactions may be represented by
\[
\sum_I {\nu '}_{I}^a\,I\; \mathrel{\mathop{\rightleftarrows}^{a _{(1)}}_{a _{(2)}}}  \; \sum_I{\nu ''}^a_{I}\, I, \quad a=1,...,r,
\]
where $a_{(1)}$ and $a_{(2)}$ are the forward and backward reactions associated to the reaction $a$, and ${\nu''}^a _{I}$, ${\nu '}^a_I$ are the forward and backward stoichiometric coefficients for the component $I$ in the reaction $a$. Mass conservation during each reaction is given by 
\[
\sum_I m_I\nu^a_{I}=0 \quad \text{for $a=1,...,r$ (Lavoisier law)},
\]
where $\nu^a _{I}:= {\nu ''}^a_{I}- {\nu'}^a _{I}$ and $m_I$ is the molecular mass of species $I$. The affinity of reaction $a$ is the state function defined by $\mathcal{A} ^a= - \sum_{I=1}^{R} \nu^a _{I} \mu ^I$, $a=1,...,r$, where $\mu^I$ is the chemical potential of the chemical species $I$. The thermodynamic flux associated with reaction $a$ is the rate of extent denoted $J_a$.

The thermodynamic displacements are $W^I$ and $\nu^a$, such that
\begin{equation}\label{therm_displ_chemistry}
\dot W^I=\mu^I,\;\; I=1,...,R\;\;\;\text{and}\;\;\; \dot\nu^a=-\mathcal{A}^a,\;\;a=1,...,r.
\end{equation}

For chemical reactions in a single compartment assumed to be adiabatically closed and without mechanical components, the variational formulation is given as follows.

\begin{framed}
\noindent Find the curves $S(t)$, $N_I(t)$, $W^I(t)$, $\nu^a(t)$, $I=1,...,R$, $a=1,...,r$, which are critical for the variational condition
\begin{equation}\label{al_GLdA_thermo_chem} 
\delta \int_{ t _1 }^{ t _2 }   \Big[L(N_1,...,N_R , S)+\dot{W}^{I}N_{I} \Big]{\rm d}t  =0,
\end{equation}
subject to the \textit{phenomenological and chemical constraints}
\begin{equation}\label{al_GNonholonomic_Constraints1_chem} 
\frac{\partial L}{\partial S}\dot S   = J_a \dot{\nu}^a\quad\text{and}\quad\dot{\nu}^a= \nu^a_I \dot{W}^I,\quad a=1,...,r,
\end{equation}
and for variations subject to the \textit{variational constraints}
\begin{equation}\label{al_GVariational_Constraints_chem} 
\frac{\partial L}{\partial S}\delta S= J_a \delta{\nu}^a \quad\text{and}\quad \delta{\nu}^a= \nu^a_I \delta W^I,\quad a=1,...,r,
\end{equation}
with $ \delta W ^I ( t _1)=\delta W^I(t_2)=0$, $I=1,...,R$.
\end{framed}

The variational formulation \eqref{al_GLdA_thermo_chem}--\eqref{al_GVariational_Constraints_chem} yields the evolution equations for chemical reactions
\[
\dot N _I= J_{a} \nu^a _{I},\quad I=1,...,R \;\;\;\text{and}\;\;\; T\dot S=J_a  \mathcal{A} ^a,
\]
together with the conditions \eqref{therm_displ_chemistry}.

Chemical reactions can be included in all the thermodynamic systems considered previously by combining the variational formulation \eqref{al_GLdA_thermo_chem}--\eqref{al_GVariational_Constraints_chem} for chemical reactions with the variational formulations \eqref{VCond_simple_diffusion}--\eqref{VC_simple_diffusion}, \eqref{VCond_conduction_matter}--\eqref{VC_conduction_matter}, and \eqref{VCond_open}--\eqref{VC_open}.}
\end{remark}

\begin{remark}[General structure of the variational formulation for open systems]{\rm As opposed to the adiabatically closed case, the phenomenological and variational constraints depend explicitly on time $t \in \mathbb{R}$ for the case of open systems. In addition, the phenomenological constraint involves an affine term that depends only on the properties at the ports. From a general point of view, letting $ \mathcal{Q}$ be the configuration manifold, these constraints are defined by the maps $A^\alpha :\mathbb{R}\times T\mathcal{Q}\rightarrow T^*\mathcal{Q}$, $A(t,x,v)\in T^*_x\mathcal{Q}$, with $A^\alpha(t,x,v)\in T^*_x\mathcal{Q}$, and $B^\alpha:\mathbb{R}\times T\mathcal{Q}\rightarrow\mathbb{R}$, $\alpha=1,...,k$, where $t \in \mathbb{R}$ and $(x,v) \in T\mathcal{Q}$.

Given a time dependent Lagrangian $\mathcal{L}:\mathbb{R}\times T\mathcal{Q}\rightarrow\mathbb{R}$ and
an external force $\mathcal{F}^{\rm ext}:\mathbb{R}\times T\mathcal{Q}\rightarrow T^*\mathcal{Q}$, the variational formulation \eqref{VCond_general}--\eqref{VC_general} is extended as follows.

\begin{equation}\label{VCond_general_open}
\delta  \int_{t_1 }^{t_2 }\mathcal{L}(t,x(t), \dot x(t)) {\rm d}t +   \int_{t_1 }^{t_2 } \left<\mathcal{F}^{\rm ext}(t,x(t), \dot x(t)), \delta x(t) \right>{\rm d}t=0, 
\end{equation}
where the curve $x(t)$ satisfies the \textit{phenomenological constraint}
\begin{equation}\label{PC_general_open}
A^\alpha(t,x,\dot x)\!\cdot\!\dot x + B^\alpha(t,x,\dot x)=0,\;\;\text{for $\alpha=1,...,k$.}
\end{equation}
and for variations $\delta x$ subject to the \textit{variational constraint}
\begin{equation}\label{VC_general_open}
A^\alpha(t,x,\dot x)\!\cdot\!\delta x =0 ,\;\;\text{for $\alpha=1,...,k$.}
\end{equation}
with $\delta x(t_1)=\delta x(t_2)=0$.

This yields the system of equations
\begin{equation}\label{general_system_open}
\left\{
\begin{array}{l}
\displaystyle\vspace{0.2cm}\frac{d}{dt} \frac{\partial \mathcal{L}}{\partial \dot x}- \frac{\partial \mathcal{L}}{\partial x} - \mathcal{F}^{\rm ext}=\lambda_\alpha A^\alpha(t,x, \dot x)\\
\displaystyle A^\alpha(t,x,\dot x)\!\cdot\!\dot x + B^\alpha(t,x,\dot x)=0,\;\;\alpha=1,...,k.
\end{array}\right.
\end{equation}
The variational formulation for open system falls into this category, by appropriate choice for $x$ and $\mathcal{L}$. For instance, for \eqref{VCond_open}--\eqref{VC_open} one has $x=(q,S,N,W, \Gamma, \Sigma)$ and $\mathcal{L}$ is the integrand in \eqref{VCond_open}. Note that in \eqref{VCond_general_open} we choose the Lagrangian to be time dependent for the sake of generality. In fact all the variational formulations for thermodynamics presented above generalize easily to time dependent Lagrangians. We refer to \cite{GBYo2018a} for a full treatment.

The energy defined by $\mathcal{E}(t,x,v)=\left<\frac{\partial \mathcal{L}}{\partial v}, v \right>- \mathcal{L}(t,x,v)$ satisfies the energy balance equation
\begin{equation}\label{energy_t}
\frac{d}{dt}\mathcal{E}=\left<\mathcal{F}^{\rm ext},\,\dot x \right>  -\lambda_\alpha B^\alpha -\frac{\partial \mathcal{L}}{\partial t}.
\end{equation}
In the application to open thermodynamic systems, the first term is identified with $P^{\rm ext}_W$, the second term is identified with $P_{H+M}^{\rm ext}$, while the third term is due to the explicit dependence of the Lagrangian on the time.}
\end{remark}

\section{Variational formulation for continuum thermodynamic systems}\label{sec_continuum}

In this section we extend Hamilton's principle of continuum mechanics \eqref{HP_continuum} to nonequilibrium continuum thermodynamics, in the same way as Hamilton's principle of classical mechanics \eqref{HP} was extended to the finite dimensional case of discrete thermodynamic systems in \S\ref{sec_finite}.

\medskip

We consider a multicomponent compressible fluid subject to the irreversible processes of viscosity, heat conduction, and diffusion. In presence of irreversible processes, we impose no-slip boundary conditions, hence the configuration manifold for the fluid motion is the manifold $Q=\operatorname{Diff}_0(\mathcal{D})$ of diffeomorphisms that keep the boundary $\partial\mathcal{D}$ pointwise fixed.

We assume that the fluid has $P$ components with mass densities $\varrho_A(t,X)$, $A=1,...,P$ in the material description, and denote by $S(t,X)$ the entropy density in the material description. The motion of the multicomponent fluid is given as before by a curve of diffeomorphisms $\varphi_t\in \operatorname{Diff}_0(\mathcal{D})$ but now $\dot\varphi_t$ is interpreted as the barycentric material velocity of the multicomponent fluid.  The Lagrangian of the multicomponent fluid with irreversible processes is
\[
L: T\operatorname{Diff}_0(\mathcal{D})\times \mathcal{F}(\mathcal{D})\times \mathcal{F}(\mathcal{D})^P\rightarrow\mathbb{R},\quad (\varphi, \dot \varphi, S, \varrho_1,...,\varrho_P) \mapsto L(\varphi, \dot \varphi, S, \varrho_1,...,\varrho_P),
\]
where $\mathcal{F}(\mathcal{D})$ denote a space of functions on $\mathcal{D}$, and is given by
\begin{equation}\label{L_fluid_multi}
\begin{aligned}
L(\varphi, \dot \varphi, S, \varrho_1,...,\varrho_P)&=K(\varphi, \dot{\varphi},\varrho_1,...,\varrho_P)- U(\varphi,S,\varrho_1,...,\varrho_P)\\
&=\int_\mathcal{D} \left[ \frac{1}{2} \varrho(X) |\dot\varphi(X)|^2 - \mathscr{E}\big(\varrho_1(X),...,\varrho_P(X),S(X),\nabla\varphi(X)\big)\right]{\rm d}^3X.
\end{aligned}
\end{equation}
The first term is the total kinetic energy of the fluid, where $\varrho:=\sum_{A=1}^P\varrho_A$ is the total mass density. The second term is minus the total internal energy of the fluid, with $\mathscr{E}$ a general expression for the internal energy density written in terms of $\varrho_A(X)$, $S(X)$, and the deformation gradient $\nabla \varphi(X)$. As in \eqref{Material_covariance} $\mathscr{E}$ satisfies the material covariance assumption and depends on the deformation gradient only through the Jacobian $J_{\varphi}$. As in \eqref{mat_cov}, there is a function $\epsilon$, the internal energy density in the spatial representation, such that
\begin{equation}\label{mat_cov_multi}
\mathscr{E}\big( \varrho_1,...,\varrho_P,\nabla \varphi)\big)=\varphi^*\big[\epsilon (\rho_1,...,\rho_P,s) \big],\quad \text{for}\quad \rho_A=\varphi_*\varrho_A,\quad s=\varphi_*S.
\end{equation}
In the spatial description, the Lagrangian \eqref{L_fluid_multi} reads
\[
\ell(\mathbf{v},s,\rho_1,...,\rho_P)=  \int_\mathcal{D}\left[\frac{1}{2}\rho |\mathbf{v}|^2 - \varepsilon (\rho_1,...,\rho_P,s) \right]{\rm d}^3x.
\]
Note that in absence of irreversible process, the Lagrangian \eqref{L_fluid_multi} would just be defined on the tangent bundle $T\operatorname{Diff}(\mathcal{D})$ with $\varrho_A=\varrho_{{\rm ref} A}$, $A=1,...,P$ and $S=S_{\rm ref}$ seen as fixed parameters, exactly as in \eqref{L_fluid} for the single component case.

\begin{remark}[Material VS spatial variational principle]{\rm As we present below, the variational formulation for continuum thermodynamical systems in the \textit{material description} is the natural continuum (infinite dimensional) version of that of discrete (finite dimensional) thermodynamical systems described in \S\ref{sec_finite}. This is in analogy with the conservative reversible case recalled earlier, namely, the Hamilton principle \eqref{HP_continuum}, associated to the material description of continuum systems, is the natural continuum version of the classical Hamilton principle \eqref{HP}. This is why we shall first consider below in \S\ref{Variational_Material} the variational formulation of continuum systems in the material description, and \textit{deduce} from it the variational formulation in the spatial description later in \S\ref{Variational_Spatial}. The latter is more involved since it contains additional constraints, as we have seen in conservative reversible case in \S\ref{sec_reduction}.}
\end{remark}

\subsection{Variational formulation in the Lagrangian description}\label{Variational_Material}

The variational formulation of a multicomponent fluid subject to the irreversible processes of viscosity, heat conduction, and diffusion, is the continuum version of the variational formulation \eqref{VCond_conduction_matter}--\eqref{VC_conduction_matter} for finite dimensional thermodynamic systems with friction, heat and mass transfer.
The analogues to the thermodynamic fluxes $F^{\rm fr}$, $J_{AB}$, $\mathcal{J}^{B\rightarrow A}$ are the viscous stress, the entropy flux density, and the diffusive flux density given by $\mathbf{P}^{\rm fr}$, $\mathbf{J}_S$, $\mathbf{J}_A$ in the material description. Total mass conservation imposes the condition $\sum_{A=1}^P\mathbf{J}_A=0$.

We give below the variational formulation for a general Lagrangian with density $\mathscr{L}$, i.e.,
\begin{equation}\label{Mat_Lagr_multi}
L(\varphi, \dot \varphi, S, \varrho_1,...,\varrho_P)=\int_\mathcal{D}\mathscr{L}\big(\varphi, \dot \varphi, \nabla\varphi, S, \varrho_1,...,\varrho_P\big){\rm d}^3X.
\end{equation}
The continuum version of the variational formulation \eqref{VCond_conduction_matter}--\eqref{VC_conduction_matter} that we propose is the following.

\begin{framed}
\noindent Find the curves $\varphi(t)$, $S(t)$, $\Gamma(t)$, $\Sigma(t)$, $W^A(t)$, $\varrho_A(t)$ which are critical for the \textit{variational condition}
\begin{equation}\label{VP_fluid} 
\int_0^T \!\! \int_ \mathcal{D}  \Big[ \mathscr{L}\big(\varphi, \dot \varphi, \nabla\varphi, S,\varrho_1,...,\varrho_P\big)  +\dot W^A  \varrho _A +  \dot \Gamma(S- \Sigma )   \Big] {\rm d} ^3X{\rm d}t =0
\end{equation}
subject to the \textit{phenomenological constraint}
\begin{equation}\label{KC_fluid}
\frac{\partial  \mathscr{L} }{\partial S}\dot \Sigma= -\mathbf{P} ^{\rm fr}: \nabla\dot \varphi  + \mathbf{J} _S \cdot \nabla \dot\Gamma  +\mathbf{J}_A \cdot \nabla \dot W ^A
\end{equation}  
and for variations subject to the \textit{variational constraint}
\begin{equation}\label{VC_fluid}
\frac{\partial  \mathscr{L}}{\partial S}\delta \Sigma = - \mathbf{P} ^{\rm fr}: \nabla \delta \varphi  + \mathbf{J} _S \cdot \nabla \delta\Gamma + \mathbf{J} _A \cdot \nabla \delta  W ^A
\end{equation} 
with $\delta \varphi(t _i )= \delta\Gamma (t _i ) =\delta W^A(t_i )=0$, $i=1,2$, and with $\delta\varphi|_{\partial\mathcal{D}}=0$.
\end{framed}

Taking variations of the integral in \eqref{VP_fluid}, integrating by parts, and using $\delta \varphi(t _i )= \delta\Gamma (t _i ) =\delta W^A(t_i )=0$, $i=1,2$, and $\delta\varphi|_{\partial\mathcal{D}}=0$, it follows
\begin{align*}
&\int_{t_1}^{t_2}\!\!\int_\mathcal{D}\left[ \Big(\frac{\partial\mathscr{L}}{\partial\varphi^a}\delta\varphi^a-\frac{\partial}{\partial t} \frac{\partial\mathscr{L}}{\partial\dot\varphi^a} - \frac{\partial }{\partial A}\frac{\partial\mathscr{L}}{\partial\varphi^a_{,A}}\Big)\delta\varphi^a + \frac{\partial \mathscr{L}}{\partial S}\delta S + \Big(\frac{\partial \mathscr{L}}{\partial \varrho_A} +\dot W^A\Big)\delta  \varrho_A\right.\\
& \hspace{4.7cm}\left.\phantom{\frac{\partial\mathscr{L}}{\partial\varphi^a_{,A}}}-\dot \varrho_A \delta W^A - (\dot S-\dot\Sigma)\delta\Gamma + \dot\Gamma(\delta S - \delta \Sigma ) \right]{\rm d}^3X{\rm d}t=0.
\end{align*}
Using the variational constraint \eqref{VC_fluid} integrating by parts and collecting the terms proportional to $ \delta \varphi $, $ \delta \Gamma $, $ \delta S$, $ \delta W^A  $, and $ \delta \varrho _A $, we get
\begin{equation}\label{condition_multi}
\begin{aligned}
\delta \varphi:&\quad\frac{d}{dt}\frac{\partial \mathscr{L} }{\partial \dot \varphi }+ \operatorname{DIV}  \Big( \frac{\partial \mathscr{L}}{\partial \nabla  \varphi } +\dot  \Gamma \frac{\partial \mathscr{L}}{\partial S}^{-1}\mathbf{P} ^{\rm fr} \Big) - \frac{\partial \mathscr{L} }{\partial \varphi }=0\\
\delta \Gamma:&\quad\dot  S=\operatorname{DIV} \Big(\dot  \Gamma \frac{\partial \mathscr{L} }{\partial S}^{-1} \mathbf{J} _S\Big) + \dot \Sigma,\hspace{2.1cm} \delta S:\quad \dot\Gamma = - \frac{\partial \mathscr{L}}{\partial S},\\
\delta W_A:&\quad\dot\varrho_A = \operatorname{DIV} \Big( \dot  \Gamma \frac{\partial \mathscr{L} }{\partial S}^{-1} \mathbf{J}_A\Big) ,\hspace{2.5cm}\delta\varrho_A:\quad  \dot W ^A =- \frac{\partial \mathscr{L} }{\partial \varrho_A },
\end{aligned}
\end{equation}
together with the boundary conditions
\[
\int_{\partial\mathcal{D}}  \mathbf{P}^{\rm fr}{}^B_a \mathbf{N}_B \delta\varphi ^a{\rm d}\mathcal S=0,\qquad \int_{\partial\mathcal{D}}  \mathbf{J}_S\cdot \mathbf{N}\delta \Gamma {\rm d}\mathcal S=0,\qquad \int_{\partial\mathcal{D}}  \mathbf{J}_A\cdot \mathbf{N}\delta W^A {\rm d}\mathcal S=0,
\]
where $\mathbf{N}$ is the outward pointing unit normal vector field to $\partial\mathcal{D}$.
The first boundary term vanishes since $\delta\varphi|_{\partial\mathcal{D}}=0$ from the no-slip boundary condition. The second and third conditions give
\[
\mathbf{J}_S\cdot \mathbf{N}=0\qquad\text{and}\qquad \mathbf{J}_A\cdot \mathbf{N}=0,\;\;A=1,...,P,\quad\text{on}\quad \partial\mathcal{D},
\]
i.e., the fluid is adiabatically closed.

From the third and fifth conditions in \eqref{condition_multi}, we have $ \dot \Gamma = - \frac{\partial \mathscr{L} }{\partial S}=\mathfrak{T} $, the temperature in material representation, and $\dot W^A =- \frac{\partial \mathscr{L} }{\partial \varrho_A }= \Upsilon^A $, a generalization of the chemical potential of component $A$ in material representation.  The second equation in \eqref{condition_multi} thus reads $\dot S+ \operatorname{DIV} \mathbf{J} _S=\dot \Sigma$ and attributes to $ \Sigma $ the meaning of \textit{entropy generation rate density}. From the first and fourth equation and the constraint, we get the system
\begin{equation}\label{multi_material} 
\left\{
\begin{array}{l}
\vspace{0.2cm}\displaystyle\frac{d}{dt}\frac{\partial \mathscr{L} }{\partial \dot \varphi }+ \operatorname{DIV}  \Big( \frac{\partial \mathscr{L} }{\partial \nabla  \varphi } -\mathbf{P} ^{\rm fr} \Big) - \frac{\partial \mathscr{L} }{\partial \varphi }=0\\
\vspace{0.2cm}\displaystyle\dot \varrho _A + \operatorname{DIV} \mathbf{J} _A=0,\quad A=1,...,P\\
\mathfrak{T} (\dot S + \operatorname{DIV} \mathbf{J} _S ) = \mathbf{P} ^{\rm fr}: \nabla\dot \varphi  - \mathbf{J} _S \cdot \nabla \mathfrak{T}  - \mathbf{J} _A \cdot \nabla \Upsilon ^A,
\end{array}
\right.
\end{equation}
for the fields $ \varphi (t, X )$, $ \varrho _A (t, X )$, and $ S(t, X)$.
The parameterization of the thermodynamic fluxes $ \mathbf{P} ^{\rm fr}$, $ \mathbf{J} _S$, $ \mathbf{J} _A $  in terms of the thermodynamic forces are discussed in the Eulerian description below.

\subsection{Variational formulation in the Eulerian description}\label{Variational_Spatial}

While the variational formulation is simpler in the material description, the resulting equations of motion are usually written and studied in the spatial description. It is therefore useful to have an Eulerian version of the variational formulation \eqref{VP_fluid}--\eqref{VC_fluid}.
In order to obtain such a variational formulation, all the variables used in \eqref{VP_fluid}--\eqref{VC_fluid} must be converted to their Eulerian analogue. We have already seen the relations $s=\varphi_*S$ and $\rho_A=\varphi_*\varrho_A$ between the Eulerian and Lagrangian mass densities and entropy densities, where the pull-back notation has been defined in \eqref{pull_back}. The Eulerian quantities associated to $\Sigma$, $\Gamma$, and $W^A$ are defined as follows
\[
\sigma= \varphi_*\Sigma,\quad \gamma= \Gamma\circ\varphi^{-1},\quad w^A=W^A\circ\varphi^{-1}.
\]
The Eulerian version of the Piola-Kirchhoff viscous stress tensor $\mathbf{P}^{\rm fr}$ is the viscous stress tensor $\boldsymbol{\sigma}^{\rm fr}$ obtained via the Piola transform, see \cite{MaHu1983}, \cite{GBYo2017b}.

From the material covariance assumption, the Lagrangian \eqref{Mat_Lagr_multi} can be rewritten exclusively in terms of spatial variables as
\[
\ell\big(\mathbf{v}, s,\rho_1,...,\rho_P\big)=\int_\mathcal{D}\mathcal{L}\big(\mathbf{v}, s,\rho_1,...,\rho_P\big){\rm d}^3x,
\]
where the Lagrangian density is defined by
\[
\mathcal{L}\big(\mathbf{v}, s,\rho_1,...,\rho_P\big)= \varphi_* \left[ \mathscr{L}(\mathbf{v}\circ\varphi, \varphi^*\rho_1,...,\varphi^*\rho_P, \varphi^*s)\right].
\]

Using all the preceding relations between Lagrangian and Eulerian variables, we can rewrite the variational formulation \eqref{VP_fluid}--\eqref{VC_fluid} in the following purely Eulerian form.

\begin{framed}
\noindent Find the curves $\mathbf{v}(t)$, $s(t)$, $\gamma(t)$, $\sigma(t)$, $w^A(t)$, $\rho_A(t)$ which are critical for the \textit{variational condition}
\begin{equation}\label{VP_fluid_Euler} 
\int_0^T \!\! \int_ \mathcal{D}  \Big[ \mathcal{L}\big(\mathbf{v}, s,\rho_1,...,\rho_P\big)  + D_t w^A  \rho _A +  D_t \gamma(s- \sigma )   \Big] {\rm d} ^3x{\rm d}t =0
\end{equation}
subject to the \textit{phenomenological constraint}
\begin{equation}\label{KC_fluid_Euler}
\frac{\partial  \mathcal{L} }{\partial s}\bar D_t \sigma= -\boldsymbol{\sigma} ^{\rm fr}: \nabla\mathbf{v}  + \mathbf{j} _S \cdot \nabla D_t\gamma  +\mathbf{j}_A \cdot \nabla D_t w ^A
\end{equation}  
and for variations $\delta \mathbf{v} =\partial _t \boldsymbol{\zeta} + \mathbf{v} \cdot \nabla \boldsymbol{\zeta} - \boldsymbol{\zeta}\cdot \nabla \mathbf{v}$, $\delta\rho_A$, $\delta w^A$, $\delta s$, $\delta\sigma$, and $\delta\gamma$ subject to the \textit{variational constraint}
\begin{equation}\label{VC_fluid_Euler}
\frac{\partial  \mathcal{L} }{\partial s}\bar D_\delta \sigma = -\boldsymbol{\sigma} ^{\rm fr}: \nabla\boldsymbol{\zeta}  + \mathbf{j} _S \cdot \nabla D_\delta \gamma + \mathbf{j} _A \cdot \nabla D_\delta  w ^A
\end{equation} 
with $\boldsymbol{\zeta}(t _i )= \delta\gamma (t _i ) =\delta w^A(t_i )=0$, $i=1,2$, and with $\boldsymbol{\zeta}|_{\partial\mathcal{D}}=0$.
\end{framed}

In \eqref{VP_fluid_Euler}--\eqref{VC_fluid_Euler} we have used the notations $D_tf=\partial _t f+ \mathbf{v}\cdot\nabla f$, $\bar D _t f= \partial _t f+ \operatorname{div}(f\mathbf{v})$, $D_\delta f=\delta f+ \boldsymbol{\zeta}\cdot\nabla f$, and $\bar D _\delta f= \delta f+ \operatorname{div}(f \boldsymbol{\zeta})$ for the Lagrangian time derivatives and variations of functions and densities.

\medskip

The variational formulation \eqref{VP_fluid_Euler}--\eqref{VC_fluid_Euler} yields the system
\begin{equation}\label{Eulerian_multi} 
\left\{
\begin{array}{l}
\vspace{0.2cm}\displaystyle
( \partial _t + \pounds _ \mathbf{v} ) \frac{\partial \mathcal{L} }{\partial \mathbf{v} }= \rho _A \nabla \frac{\partial \mathcal{L} }{\partial \rho _A }+ s \nabla \frac{\partial \mathcal{L} }{\partial s }+ \operatorname{div} \boldsymbol{\sigma}  ^{\rm fr}\\
\vspace{0.2cm}\displaystyle \bar D_ t\rho _A + \operatorname{div} \mathbf{j} _A=0,\quad A=1,...,P \\
\displaystyle \frac{\partial \mathcal{L} }{\partial s } (\bar D_t s + \operatorname{div} \mathbf{j} _s) = - \boldsymbol{\sigma} ^{\rm fr} \!: \!\nabla \mathbf{v} - \mathbf{j} _s\! \cdot  \!\nabla \frac{\partial \mathcal{L} }{\partial s }- \mathbf{j} _A \!\cdot \! \nabla \frac{\partial \mathcal{L} }{\partial \rho _A },
\end{array}
\right.
\end{equation}
together with the conditions
\[
\bar D_t\sigma= \bar D_t s + \operatorname{div} \mathbf{j} _s,\quad D_t\gamma= - \frac{\partial \mathcal{L} }{\partial s }, \quad D_tw_A= - \frac{\partial \mathcal{L} }{\partial \rho_A }.
\]
In \eqref{Eulerian_multi} $\pounds_\mathbf{v}$ denotes the Lie derivative defined as $\pounds_\mathbf{v}\mathbf{m}=\mathbf{v}\cdot \nabla\mathbf{m} + \nabla \mathbf{v}^\mathsf{T}\cdot \mathbf{m} + \mathbf{m}\operatorname{div}\mathbf{v}$.
We refer to \cite{GBYo2017b} a detailed derivation of these equations from the variational formulation  \eqref{VP_fluid_Euler}--\eqref{VC_fluid_Euler}.

\paragraph{The multicomponent Navier-Stokes-Fourier equations.} For the Lagrangian
\[
\ell\big(\mathbf{v}, s,\rho_1,...,\rho_P\big)=\int_\mathcal{D}\left[\frac{1}{2}\rho|\mathbf{v}|^2 -\epsilon(\rho_1,...,\rho_P,s\big)\right]{\rm d}^3x
\]
we get
\begin{equation}\label{multi_NSF} 
\left\{
\begin{array}{l}
\vspace{0.2cm}\displaystyle
\rho ( \partial_t \mathbf{v} +\mathbf{v} \cdot \nabla \mathbf{v} )= - \nabla p +\operatorname{div} \boldsymbol{\sigma} ^{\rm fr}\\
\displaystyle\vspace{0.2cm} \bar D_ t\rho _A + \operatorname{div} \mathbf{j} _A=0,\quad A=1,...,P \\
\displaystyle T (\bar D_t s + \operatorname{div} \mathbf{j} _s) = \boldsymbol{\sigma} ^{\rm fr} : \nabla \mathbf{v} - \mathbf{j} _s \cdot  \nabla T - \mathbf{j} _A \cdot  \nabla \mu ^A 
\end{array}
\right.
\end{equation}
with $\mu^A= \frac{\partial \epsilon}{\partial\rho_A}$, $T=  \frac{\partial \epsilon}{\partial s}$, and $p= \mu^A\rho_A + Ts- \epsilon$.

\medskip

The system of equations \eqref{multi_NSF} needs to be supplemented with phenomenological expressions for the \textit{thermodynamic fluxes} $ \boldsymbol{\sigma} ^{\rm fr}$, $ \mathbf{j} _S$, $ \mathbf{j} _A $ in terms of the \textit{thermodynamic affinities} $ \operatorname{Def} \mathbf{v} $, $ \nabla T$, $ \nabla \mu ^A $ compatible with the second law. It is empirically accepted  that for a large class of irreversible processes and under a wide range of experimental conditions, the thermodynamic fluxes $ J_ \alpha $ are linear functions of the thermodynamic affinities $ X^\alpha $, i.e., $J_ \alpha = \mathcal{L} _{ \alpha \beta } X^\beta $, where the transport coefficients $ \mathcal{L} _ { \alpha \beta }(...)$ are state functions that must be determined by experiments or if possible by kinetic theory.
Besides defining a positive quadratic form, the coefficients $ \mathcal{L} _ { \alpha \beta }(...)$ must also satisfy {\it Onsager's reciprocal relations} (\cite{Onsager1931}) due to the microscopic time reversibility and the {\it Curie principle} associated to material invariance (see, for instance, \cite{deGrootMazur1969}, \cite{KoPr1998}, \cite{LaLi1969_6}, \cite{Woods1975}). 
In the case of the multicomponent fluid, writing the traceless part of $\boldsymbol{\sigma}^{\rm fr}$ and $\operatorname{Def} \mathbf{v}$ as $ (\boldsymbol{\sigma} ^{\rm fr})^{(0)}=\boldsymbol{\sigma} ^{\rm fr}- \frac{1}{3}( \operatorname{Tr}\boldsymbol{\sigma} ^{\rm fr} )\delta $ and $ (\operatorname{Def} \mathbf{v}) ^{(0)}=  \operatorname{Def} \mathbf{v} - \frac{1}{3}(\operatorname{div} \mathbf{v} )\delta $, we have the following phenomenological linear relations
\[
-\begin{bmatrix}
\mathbf{j} _S\\
\mathbf{j} _A
\end{bmatrix}= 
\begin{bmatrix}
\mathcal{L} _{SS} & \mathcal{L} _{SB}\\
\mathcal{L} _{AS}&\mathcal{L} _{AB}
\end{bmatrix}\begin{bmatrix}
\nabla T\\
\nabla \mu ^B 
\end{bmatrix}, \qquad 
\frac{1}{3}\operatorname{Tr}\boldsymbol{\sigma} ^{\rm fr}= \zeta \operatorname{div} \mathbf{v}, \qquad
(\boldsymbol{\sigma} ^{\rm fr}) ^{(0)}= 2 \mu( \operatorname{Def}\mathbf{v})^{(0)}\!,
\]
where all the coefficients may depend on $(s, \rho_1,...,\rho_P)$. The first linear relation describes the vectorial phenomena of heat conduction (Fourier law), diffusion (Fick law) and their cross effects (Soret and Dufour effects), the second relation describes the scalar processes of bulk viscosity with coefficient $\zeta\geq 0$, and the third relation is the tensorial process of shear viscosity with coefficient $\mu\geq 0$.
The associated friction stress reads
\[
\boldsymbol{\sigma} ^{\rm fr}= 2 \mu \operatorname{Def} \mathbf{v} + \Big( \zeta - \frac{2}{3} \mu \Big) (\operatorname{div} \mathbf{v} ) \delta.
\]

All these phenomenological considerations take place in the phenomenological constraint \eqref{KC_fluid_Euler} and the associated variational constraints \eqref{VC_fluid_Euler}, but they are not involved in the variational condition \eqref{VP_fluid_Euler}. Note that our variational formulation holds independently on the linear character of the phenomenological laws.

\begin{remark}{\rm For simplicity, we chose the fluid domain $\mathcal{D}$ as a subset of $\mathbb{R}^3$ endowed with the Euclidean metric. More generally, the variational formulation can be intrinsically written on Riemannian manifolds, see \cite{GBYo2017b}. Making the dependence of the Riemannian metric explicit, even if it is given by the standard Euclidean metric, is important for the study of the covariance properties, \cite{GBMaRa2012}.}
\end{remark}

\section{Concluding remarks}\label{conclusions}
In this paper, we made a  survey of our recent developments on the Lagrangian variational formulation for nonequilibrium thermodynamics developed in \cite{GBYo2017a, GBYo2017b, GBYo2018a}, which is a natural extension of Hamilton's principle in mechanics to include irreversible processes.

Before going into details, we made a brief review of Hamilton's principle as it applies to (finite dimensional) discrete systems in classical mechanics as well as to (infinite dimensional) continuum systems. Then, in order to illustrate our variational formulation for nonequilibrium thermodynamics, we first started with the finite dimensional case of adiabatically closed systems together with representative examples such as a piston containing an ideal gas, a system with a chemical species experiencing diffusions between several compartments, an adiabatic piston with two cylinders, and a system with a chemical species experiencing diffusion and heat conduction between two compartments. Then, we extended the variational formulation to open finite dimensional systems that can exchange heat and matter with the exterior. This case was illustrated with the help of a piston device with ports and heat sources. We also demonstrated how chemical reactions can be naturally incorporated into our variational formulation.

Second, we illustrated the variational formulation with the infinite dimensional case of continuum systems by focusing on a compressible fluid with the irreversible processes due to viscosity, heat conduction, and diffusion. The formulation is first given in the Lagrangian (or material) description because it is in this description that the variational formulation is a natural continuum extension of the one for discrete systems. The variational formulation in the Eulerian (or spatial) description is then deduced by Lagrangian reduction and yields the multicomponent Navier-Stokes-Fourier equations.

One of the key issue of our variational formulation is the introduction and the use of the concept of thermodynamic displacement, whose time derivative corresponds to the affinity of the process. The thermodynamic displacement allows to systematically develop the variational constraints associated to the nonlinear phenomenological constraints. The variational formulations presented in this paper use the entropy as an independent variable, but a variational approach based on the temperature can be also developed by considering free energy Lagrangians, see \cite{GBYo2018b}.

\paragraph{Further developments.} Associated with our variational formulation of nonequilibrium thermodynamics, there are the following interesting and important topics, which we have not described here due to lack of space but are quite relevant with the variational formulation of nonequilibrium thermodynamics, reviewed in this paper.
\begin{itemize}
\item \textbf{Dirac structures and Dirac  systems:} It is well-known that when the Lagrangian is regular, the equations of classical mechanics can be transformed into the setting of Hamiltonian systems. The underlying geometric object for this formulation is the canonical symplectic form on the phase space $T^*Q$ of the configuration manifold. When irreversible processes are included, this geometric formulation is lost because of the degeneracy of the Lagrangians and the presence of the nonlinear nonholonomic constraints. Hence one may ask what is the appropriate geometric object that generalizes the canonical symplectic form in the formulation of thermodynamics. In \cite{GBYo2018c,GBYo2018e} it was shown that the evolution equations for both adiabatically closed and open systems can be geometrically formulated in terms of various classes of Dirac structures induced from the phenomenological constraint and from the canonical symplectic form on $T ^\ast Q$ or on $T ^\ast (Q\times \mathbb{R}  )$.
\item \textbf{Reduction by symmetry:} When symmetries are available, reduction processes can be applied to the variational formulation of thermodynamics, thereby extending the process of Lagrangian reduction from classical mechanics to thermodynamics. This has been already illustrated in \S\ref{Variational_Spatial} for the Navier-Stokes-Fourier equation, but can be carried out in general for all the variational formulations presented in this paper. For instance, we refer to \cite{CoGB2018} for the case of simple thermodynamic systems on Lie groups with symmetries.
\item \textbf{Variational discretization:} Associated to the variational formulation in this paper, there exist variational integrators for the nonequilibrium thermodynamics of simple adiabatically closed systems, see \cite{GBYo2018d}, \cite{CoGB2018}. These integrators are structure preserving numerical schemes that are obtained by a discretization of the variational formulation. The structure preserving property of the flow of such systems  is an extension of the symplectic property of the flow of variational integrators for Lagrangian mechanics.
\item \textbf{Modelling of thermodynamically consistent models:} The variational formulation for thermodynamics can be also used to derive new models, which are automatically thermodynamically consistent. We refer to \cite{GB2018} for an application of the variational formulation to atmospheric thermodynamics and its pseudoincompressible approximation.
\end{itemize}


\paragraph{Acknowledgments.} F.G.B. is partially supported by the ANR project GEOMFLUID, ANR-14-CE23-0002-01; H.Y. is partially supported by JSPS Grant-in-Aid for Scientific Research (16KT0024, 24224004), the MEXT ``Top Global University Project'' and Waseda University (SR 2018K-195).

\end{document}